\documentclass[aps,pra,reprint,twocolumn,superscriptaddress,nofootinbib]{revtex4-2}
\pdfoutput=1 
\usepackage{graphicx}
\usepackage{bm}
\usepackage{amsmath}
\usepackage{amssymb}
\usepackage{amsfonts}
\usepackage{amsthm}
\usepackage{mathtools}
\usepackage{hyperref}
\usepackage{braket}
\usepackage[normalem]{ulem}
\usepackage{wrapfig}
\usepackage{tikz}
\usepackage{dsfont}
\usepackage{comment}

\hypersetup{colorlinks=true,urlcolor=[rgb]{0,0,0.5},citecolor=[rgb]{0.5,0,0},linkcolor=[rgb]{0,0,0.4}}

\def\pPsi{\Psi}

\def\beq{\begin{equation}}
\def\eeq{\end{equation}}

\begin{document}
\title{Large $N$ fractons}

\author{Kristan Jensen}
\email{kristanj@uvic.ca}
\affiliation{\it Department of Physics and Astronomy, University of Victoria, Victoria, BC V8W 3P6, Canada}
\author{Amir Raz}
\email{araz@utexas.edu}
\affiliation{\it University of Texas, Austin, Physics Department, Austin TX 78712, USA}

\begin{abstract}
We consider theories of fractons with $N$ fields. These theories have exotic spacetime symmetries, including a conserved dipole moment. Using collective fields we solve these models to leading order in large $N$. The large $N$ solution reveals that these models are strongly correlated, and that interactions generate a momentum-dependent self-energy. Dipole symmetry is spontaneously broken throughout the phase diagram of these models, leading to a low-energy Goldstone description of what we dub ``dipole superfluids.''
\end{abstract}

\maketitle

%--------------------------------------
\textit{Introduction.}~Consider quantum mechanical models of fracton order (see e.g.~\cite{Chamon:2004lew, 2011AnPhy.326..839B, Haah:2011drr, Vijay:2015mka,  Nandkishore:2018sel, GromovMultipole}). These models describe new, at this time purely hypothetical phases of quantum matter. Perhaps the two most visceral signatures of these phases are finite-energy quasiparticles of restricted mobility, colloquially called ``fractons,'' and in some cases, a large ground state degeneracy sensitive to the details of an underlying lattice. While these phases have yet to be realized in experiment\footnote{Although perhaps they have~\cite{PhysRevX.10.011042}.} there is reason to believe that they will be found. Besides the prospect of engineering such a phase with ultra-cold atoms~\cite{sous2019} or antiferromagnets~\cite{Sous:2020ypq}, there are proposals that defects in elastic media in $2+1$-dimensions~\cite{2018PhRvL.120s5301P, GromovElasticity}, vortices in superfluid helium~\cite{Nguyen:2020yve,Doshi:2020jso}, and the lowest Landau level of fractional quantum Hall states~\cite{Du:2021pbc} all exhibit fracton order.

Fractons pose a number of challenges to condensed matter and high energy theorists. From a purely theoretical point of view, one ought to be able to obtain a continuum field theory description of fracton phases upon coarse-graining, but then the features mentioned above seem to defy the standard Wilsonian effective field theory paradigm. Informally, this challenge can be phrased as a question: how can field theory describe restricted-mobility excitations or a UV-sensitive spectrum of low-energy states? (See~\cite{Seiberg:2020bhn,Seiberg:2020wsg}.) There is another practical problem for theorists to solve. Apart from completely integrable Hamiltonians like the X-Cube model~\cite{Vijay:2016phm} or extreme limits of condensed phases, generic models of fractons are strongly correlated, and as such we have few tools to study them and little knowledge of their macroscopic features, which if they were known could be used to find them in Nature. Can we find soluble models of interacting fractons, and thereby study their physics?

The goal of this Letter is to address these theoretical and practical challenges. An important clue is that, in understood examples, the existence of restricted-mobility excitations and a lattice-sensitive ground state degeneracy are ultimately consequences of exotic spacetime symmetries. In the experimental proposals for fracton order above, like vortices in superfluid helium, the symmetry generators include both a $U(1)$ charge and, crucially, the corresponding dipole moment.\footnote{This example also include a conserved quadrupole trace.} Isolated charges are then immobile by symmetry; these are the sought-after fractons. Charges can bind into completely mobile dipoles. Meanwhile, if a ground state is not invariant under the dipole symmetry, there is a ground state manifold generated by action of the dipole symmetry, with a ground state degeneracy parametrically equal to the volume of the dipole symmetry group. This volume is infinite in the continuum, but it is compact in a lattice regularization.  

This clue is important because it suggests a way forward, which we take in this Letter. Namely, we find and solve simple theories with these exotic spacetime symmetries, paying careful attention to subtleties that arise in the functional integral. 

Inspired by models of Pretko~\cite{Pretko_2018}, we study interacting continuum field theories with $N$ charged scalar fields, imposing a $U(N)$ symmetry that rotates the fields, and a conserved dipole moment associated with the diagonal $U(1)\subset U(N)$ charge. The Lagrangians of these theories are non-standard. Instead of the usual quadratic terms with two spatial derivatives, the simplest terms with spatial derivatives include at last four powers of the fundamental fields and these may lead to strong interactions. However, these theories are generalized vector models, which we proceed to solve in the large $N$ limit using methods familiar from the large $N$ Chern-Simons-matter and SYK literature (e.g.~\cite{Giombi:2011kc,Rosenhaus:2018dtp}).

We establish many results for these theories, including their phase diagram at finite temperature and chemical potential. Crucially the dipole symmetry is spontaneously broken everywhere in the phase diagram, and we find a new high temperature phase in which the dipole symmetry is spontaneously broken, but the $U(1)$ symmetry is not. From a more theoretical viewpoint, the interactions mentioned above with spatial derivatives generate a momentum-dependent self-energy, which tames loop integrals. When there is an underlying lattice the large $N$ solution has a continuum limit, but the spectrum of fluctuations retains some sensitivity to the details of the lattice. For example, the zero modes associated with the symmetry breaking have a UV-sensitive volume, the number of lattice sites in a lattice regularization, which allows the near-continuum theory to describe a UV-sensitive spectrum of low-energy states. 

The remainder of this Letter is organized as follows. We write down the the large $N$ models of interest in the next Section, and rewrite them in terms of collective fields that are weakly coupled at large $N$. We then find the solution to the collective field description and the ensuing phase diagram, and conclude with a Discussion. Many further results are relegated to the Appendix.

%--------------------------------------
\textit{Models with conserved dipole moment.}~We work at finite temperature through the imaginary time formalism. We consider models with $N$ complex scalars $\phi^a$ (with $a=1,2,\hdots,N$) enjoying a number of symmetries. To wit, we impose invariance under spacetime translation symmetry, spatial rotations, parity, and $U(N)$ symmetry. Crucially, we also demand that the dipole moment associated with the diagonal $U(1)\subset U(N)$ is conserved. The latter amounts to an invariance under $\phi^a(t,\vec{x}) \to e^{i \vec{d} \cdot \vec{x}}\phi^a(\tau,\vec{x})$, which we term a dipole transformation. Pretko~\cite{Pretko_2018} has found a useful way to write down effective actions invariant under dipole transformations. While spatial derivatives of $\phi^a$ are not covariant under dipole transformations, the basic covariant object with spatial derivatives is
\beq 
	D_{ij}(\phi^a,\phi^b) = \frac{1}{2}\left( \phi^a \partial_i \partial_j \phi^b - \partial_i \phi^a \partial_j \phi^b + (a\leftrightarrow b)\right)\,,
\eeq
which transforms as $D_{ij}(\phi^a,\phi^b) \to e^{2i\vec{d}\cdot \vec{x}}D_{ij}(\phi^a,\phi^b)$.

In this work we study simple field theories with a single time derivative, at most quartic interactions, and at most four spatial derivatives. There is a model with only two spatial derivatives, which we call \underline{Model 1}, given by
\beq \label{eq:action_model_1}
	S = \int d\tau d^dx \left( \bar{\phi}^a\partial_{\tau} \phi^a + 2\text{Re}\left[ \frac{\lambda}{N} \delta^{ij}D_{ij}(\bar{\phi}^a,\bar{\phi}^b) \phi^a \phi^b\right] + V\right)\,,
\eeq
where $V =- \mu \bar{\phi}^a\phi^a +\frac{ \lambda_4}{N} (\bar{\phi}^a\phi^a)^2$, and sums over $a,b$ are implied. The model is specified by a chemical potential $\mu$, an inverse temperature $\beta$ (introduced by analytic continuation to imaginary time), a quartic interaction $\lambda_4$, and a complex coupling $\lambda$. We have introduced factors of $1/N$ so that there is a nice large $N$ limit. 

We also introduce \underline{Model 2}, which has four spatial derivatives, described by
\begin{align} 
	\nonumber
	S =& \int d\tau d^dx \left( \bar{\phi}^a\partial_{\tau} \phi^a + \frac{\lambda_T}{N} D_{\{ij\}}(\phi^a,\phi^b)D^{\{ij\}}(\bar{\phi}^a,\bar{\phi}^b)\right.
	\label{eq:action_model_2}
	\\
	& \qquad \left. + \frac{\lambda_S}{N}\ \left| \delta^{ij}D_{ij}(\phi^a,\phi^b)+ \gamma \phi^a \phi^b\right|^2 + V\right)\,,
\end{align}
where $D_{\{ij\}}(\phi^a,\phi^b) = D_{ij}(\phi^a,\phi^b) - \frac{\delta_{ij}}{d}\delta^{kl}D_{kl}(\phi^a,\phi^b)$ is the traceless part of $D_{ij}(\phi^a,\phi^b)$. In addition to the chemical potential and temperature, this model is characterized by real couplings $\lambda_T$ ($T$ is for tensor) and $\lambda_S$ ($S$ is for scalar), and a complex parameter $\gamma$. Note that one can reach Model 1 by a scaling limit of Model 2, by taking $\lambda_T ,\lambda_S\to 0$ while holding $\lambda_S \gamma = \lambda$ fixed.

Both Model 1 and Model 2 are effectively vector models, and so are soluble at large $N$. To solve them, we integrate in bilocal collective degrees of freedom $G(x_1,x_2)$ and $\Sigma(x_1,x_2)$, whose expectation values are the large $N$ propagator of $\phi^a$ and self-energy respectively. These fields decouple the quartic interactions, allowing us to integrate out the $\phi^a$. We then integrate all but one of the scalars, $\phi^1 = \sigma$, which will allow us to diagnose the spontaneous breaking of $U(N) \to U(N-1)$. The fields $(G,\Sigma;\sigma)$ are weakly coupled at large $N$, and solving the large $N$ model amounts to solving their classical equations of motion. Making a translationally invariant ansatz, with $\sigma(x) = \sigma$ and
\beq
	G(k_1,k_2) = NG(k_2) \Delta \,,  \quad \Sigma(k_1,k_2) = \Sigma(k_2) \Delta\,,
\eeq
where we have Fourier transformed and $\Delta = \beta \delta_{n_1n_2} (2\pi)^d \delta(\vec{k}_1+\vec{k}_2)$ with $\omega_n = \frac{2\pi n}{\beta}$ the frequency of the $n^{\rm th}$ Matsubara mode, these equations (the large $N$ Dyson equations for the propagator and self-energy) read
\begin{align}
\begin{split}
\label{E:dyson}
	G(k)  &= \frac{1}{i\omega_n + \Sigma(k)}+\frac{|\sigma|^2}{N}\beta\delta_{n0}(2\pi)^d \delta^d(\vec{k})\,,
	\\
	\Sigma(k) & =-\mu +  2 \int Dk' \,G(k') NV_4(-k,-k',k,k')\,,
\end{split}
\end{align}
along with $\sigma \Sigma(k=0) =0$. Here $Dk = \frac{1}{\beta} \sum_n \frac{d^dk}{(2\pi)^d}$ and $NV_4$ is the quartic vertex in momentum space. Model 1 has $NV_4 = \frac{1}{2}(\lambda |\vec{k}_{12}|^2 + \bar{\lambda} |\vec{k}_{34}|^2) + \lambda_4$ with $\vec{k}_{mn} = \vec{k}_m -\vec{k}_n$, which gives a local version of an $M$-particle Hamiltonian considered in~\cite{Glorioso:2021bif}. In Model 2 the quartic vertex is
\begin{align}
\begin{split}
	NV_4(k_i) &= \frac{\lambda_T}{4}\left( (\vec{k}_{12}\cdot \vec{k}_{34})^2 - |\vec{k}_{12}|^2|\vec{k}_{34}|^2\right)
	\\
	& \quad + \frac{\lambda_S}{4}(|\vec{k}_{12}|^2 + 2 \bar{\gamma})(|\vec{k}_{34}|^2 + 2\gamma) + \lambda_4\,.
\end{split}
\end{align}
Diagrammatically the Dyson equations resum bubbles as shown in Fig.~\ref{F:dyson}. A more detailed algebraic derivation is given in Appendix \ref{app:derivation}. 

\begin{figure}[t]
\includegraphics[width=1.7in]{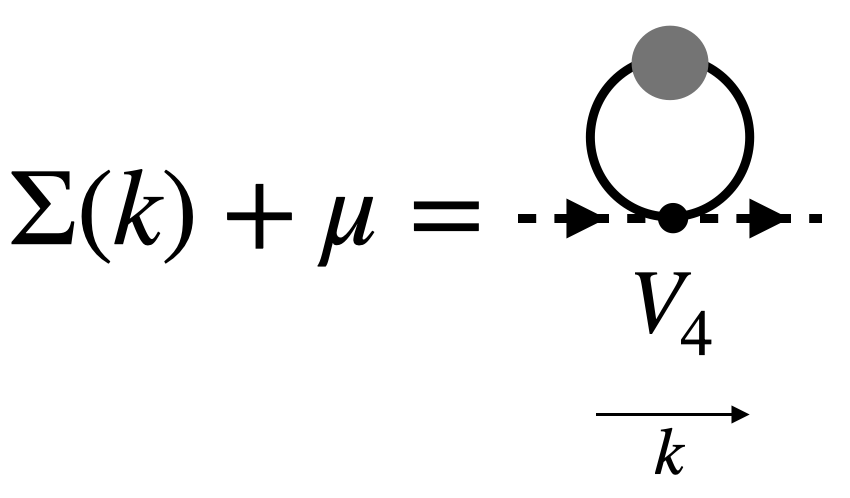}
\caption{\label{F:dyson} The Dyson equations in diagrammatic form. The internal line is the exact large $N$ propagator $G$.}
\end{figure}

These models are many-body quantum mechanics in $d$ dimensions, where a conserved number of particles $M$ (conjugate to $\mu$) interact via $2\to 2$ interactions characterized by $V_4(k_i)$, where $(k_1,k_2)$ label the incoming momenta and $(k_3,k_4)$ the outgoing momenta. The part of the potential that appears in the Dyson equation~\eqref{E:dyson}, $V_4(-k,-k',k,k')$, is the forward limit of $V_4$. In Model 1 it is $\text{Re}(\lambda)|\vec{k}-\vec{k}'|^2 + \lambda_4$, and in Model 2 it is $\frac{\lambda_S}{4} \left| |\vec{k}-\vec{k}'|^2 + 2\gamma\right|^2 + \lambda_4$. We anticipate our Models to be consistent only when the forward limit is positive. For Model 1 this implies $\text{Re}(\lambda),\lambda_4\geq 0$, with more complicated constraints for Model 2.

Let us return to the dipole symmetry. It acts on the momentum space field by $\tilde{\phi}^a(\omega,\vec{k}) \to \tilde{\phi}^a(\omega,\vec{k}+\vec{d})$, i.e. by a shift of momentum, and on the conjugate field by the opposite shift. In a lattice regularization $\vec{d}$ is then valued in the space of momenta, the Brillouin zone. Dipole transformations then act on $G(k)$ and $\Sigma(k)$ by $G(\omega,\vec{k})\to G(\omega,\vec{k}+\vec{d})$, and similarly for $\Sigma$. It follows that the Green's function is an order parameter for dipole breaking: if $G(\omega,\vec{k})$ depends on the spatial momentum, i.e. if the position space Green's function $\langle \bar{\phi}^a(x)\phi^b(0)\rangle$ is not ultralocal in space, then the dipole symmetry acts on it. This makes physical sense: $G$ is the one-point function of an operator that creates a dipole, with a charge at $0$ and an anticharge at $x$, which if nonzero is a dipole condensate in analogy with the usual charge condensate associated with ordinary symmetry breaking.

Insofar as it takes fine-tuning to make the Green's function ultralocal, we expect the dipole symmetry to be broken throughout the phase diagram. There is also the possibility of $U(N)\to U(N-1)$ symmetry breaking, parameterized by the condensate $\sigma = \langle \phi^{a=1}\rangle$, and in such a phase dipole symmetry is necessarily broken as well. That is, on general grounds we expect to find two phases of our models: (i.) a ``normal'' phase where dipole symmetry is broken but $U(N)$ is preserved,  and (ii.) a ``condensed'' phase in which $U(N)$ is broken to $U(N-1)$ and the dipole symmetry is also broken. In both phases the dipole symmetry is broken, and so both phases are a superfluid of condensed dipoles. The ``normal'' phase has a vector order parameter, $\partial_i G$ in the concident limit, so we call it a \textit{p-wave dipole superfluid}, while the ``condensed'' phase has a scalar order parameter $\langle \phi^a\rangle$, so we call it a \textit{s-wave dipole superfluid}.

In the absence of a condensate $\sigma$ the large $N$ equations~\eqref{E:dyson} are invariant under dipole shifts on account of the fact they only depend on the difference of momenta $\vec{k}-\vec{k}'$. As such, a solution $G(\omega,\vec{k})$ implies the existence of a family of solutions $G(\omega,\vec{k}+\vec{d})$. With a condensate, we get a family of solutions provided that we also shift the condensate as $\delta^d(\vec{k})\to \delta^d(\vec{k}+\vec{d})$.

%--------------------------------------
\textit{Large $N$ solutions.}~We now endeavor to solve the Dyson equations~\eqref{E:dyson}. Because the vertex $V_4$ does not depend on the frequency $\omega_n$, it follows that the self-energy $\Sigma$ only depends on spatial momentum. This allows us to perform the sum over Matsubara modes in~\eqref{E:dyson}, so that
\begin{align}
\nonumber
    \Sigma(\vec{k}) = &\int \frac{d^dk'}{(2\pi)^d}  NV_4(-k,-k',k,k') \coth\left(\frac{\beta \Sigma(\vec{k}') }{ 2}\right) 
    \\
     \label{E:Dyson2}
    &  -\mu +  2\frac{|\sigma|^2}{N} NV_4(-k,0,k,0),
\end{align}
along with $\sigma \Sigma(k=0) = 0$. 

\begin{figure*}
    \centering
    \includegraphics[width = 0.97\linewidth]{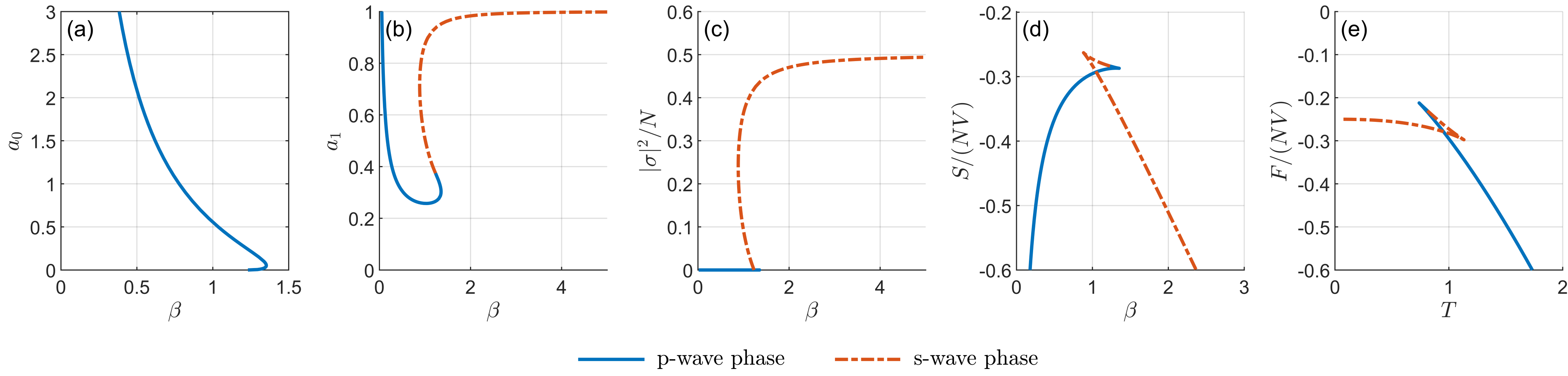}
    \caption{The solutions to the Dyson equations for Model 1 are presented in panels \textbf{(a)} and \textbf{(b)}, along with the condensate in \textbf{(c)}, the on-shell action in \textbf{(d)}, and free energy densities in \textbf{(e)} as a function of inverse temperature $\beta$  in $d=3$, with $\mu = 1$ and $\text{Re}(\lambda) = \lambda_4 = 1$.}
    \label{fig:mod1_pt}
\end{figure*}

The vertex $V_4(-k,-k',k,k')$ is a polynomial in $k$ of degree 2 in Model 1, and degree 4 in Model 2,
and so $\Sigma(\vec{k})$ is too. We then make a rotationally invariant ansatz $\Sigma(\vec{k}) = a_0 + a_1 |\vec{k}|^2+a_2 |\vec{k}|^4$ (with $a_2=0$ for Model 1). The Dyson equation \eqref{E:Dyson2} becomes a finite  system of coupled equations for the unknown quantities $a_0,a_1,a_2$ and $\sigma$. The condition $\sigma \Sigma(k=0) = \sigma a_0 = 0$ leads to two classes of solutions, a normal phase with $\sigma = 0$ and $a_0 \neq 0$, and a condensed phase with $\sigma \neq 0$. 

Plugging our ansatz back into~\eqref{E:Dyson2} and suitably regularizing, the momentum integral can be evaluated analytically in Model 1 and numerically in Model 2. Here we focus on Model 1, leaving Model 2 to the Appendix. In the normal phase where $\sigma = 0$,~\eqref{E:Dyson2} becomes
\begin{equation} \label{eq:dyson_mod1_N}
\begin{split}
    a_0 & =  -\mu +  \frac{d}{2\beta} \frac{\text{Li}_{\frac{d+2}{2}}\left( e^{-\beta a_0}\right)}{\text{Li}_{\frac{d}{2}}\left( e^{-\beta a_0}\right)}  + \frac{\lambda_4}{\text{Re}(\lambda)} a_1 , \\
	a_1 & = \frac{2 \text{Re}(\lambda)}{(4\pi \beta a_1)^{\frac{d}{2}}}\text{Li}_{\frac{d}{2}}\left(e^{-\beta a_0}\right) ,
\end{split}
\end{equation}
while in the condensed phase where $a_0 = 0$ we have
\begin{equation} \label{eq:dyson_mod1_C}
\begin{split}
     \mu & = \frac{\lambda_4}{\text{Re}(\lambda)} a_1  +  \frac{d~ \text{Re}(\lambda)}{(4\pi \beta a_1)^{\frac{d}{2}}} \frac{1}{\beta a_1} \zeta\left(\frac{d+2}{2}\right),\\
    \frac{|\sigma|^2}{N} &= \frac{a_1}{2 \text{Re}(\lambda)}  -  \frac{1}{(4\pi \beta a_1)^{\frac{d}{2}}}\zeta\left(\frac{d}{2}\right) .
\end{split}
\end{equation}

In both models we solve the Dyson equations numerically using standard solvers of systems of nonlinear equations (e.g. the function ``fsolve'' in MATLAB or ``NSolve'' in Mathematica.) We present some numerical solutions for Model 1 in Fig. \ref{fig:mod1_pt} and for Model 2 in the Appendix.

We can also solve the Dyson equations analytically in various limits of temperature and chemical potential. For example, in Model 1 at low temperature we find $a_1 \approx \text{Re}(\lambda)\mu/\lambda_4$, $|\sigma|^2/N \approx \mu/(2\lambda_4)$ when $\mu > 0$ and $d> 2$; $a_1 \approx \text{Re}(\lambda)\mu/\lambda_4$, $ a_0 \approx 0,$ and $ \sigma=0$ when $\mu > 0$ and $d= 2$; and $a_0 \approx -\mu$, $a_1^{\frac{d+2}{2}} \approx \frac{2\text{Re}(\lambda)}{(4\pi \beta)^{\frac{d}{2}}}e^{\beta \mu}$, and $\sigma=0$ when $\mu<0$. We present more asymptotic solutions in Appendix \ref{app:limits}. In fact, at zero temperature we find that the large $N$ solution is one-loop exact in $\lambda, \lambda_S,$ and $\gamma$ to leading order in large $N$, although in general the solution is a non-analytic function of couplings.

%--------------------------------------
\textit{Phase diagrams.}~Our models have, in general, a non-trivial phase diagram. To determine it we evaluate the thermal free energy density from our large $N$ solutions. We have in finite but large volume
\beq
	\beta F =-\ln Z = NS^{(0)} +S^{(1)} + \frac{S^{(2)}}{N} + O(N^{-2})\,,
\eeq
where the $1/N$ expansion is the weak coupling expansion of the collective field theory of $(G,\Sigma;\sigma)$.
Here $NS^{(0)}$ is the on-shell action of the large $N$ solution, while $S^{(1)}= -\ln Z_{\rm 1-loop}= - \ln V_{\rm 1-loop} - \ln \widetilde{Z}_{\rm 1-loop}$ is the one-loop correction which we separate into a zero mode volume $V_{\rm 1-loop} = VV_{\rm BZ}=N_{\rm sites}$ the dimensionless volume of the Brillouin zone, i.e. the number of lattice sites, and a contribution $-\ln \widetilde{Z}_{\rm 1-loop}$ from nonzero modes, $S^{(2)}$ the two-loop correction, and so on. The dipole breaking implies a UV-sensitive normalization $Z \propto N_{\rm sites}$. In the Appendix we show that this prefactor can be simply understood from the symmetry algebra. Irreducible representations of the dipole symmetry with nonzero charge have a dimension $N_{\rm sites}$, implying that the density of states has that prefactor, and so also $Z$.

In most regions of the phase diagram we only find a single solution. In those where both phases exist, the dominant one is that with the lower large $N$ free energy, i.e. smaller on-shell action.

\begin{figure}
    \centering
    \includegraphics[width = 0.7\linewidth]{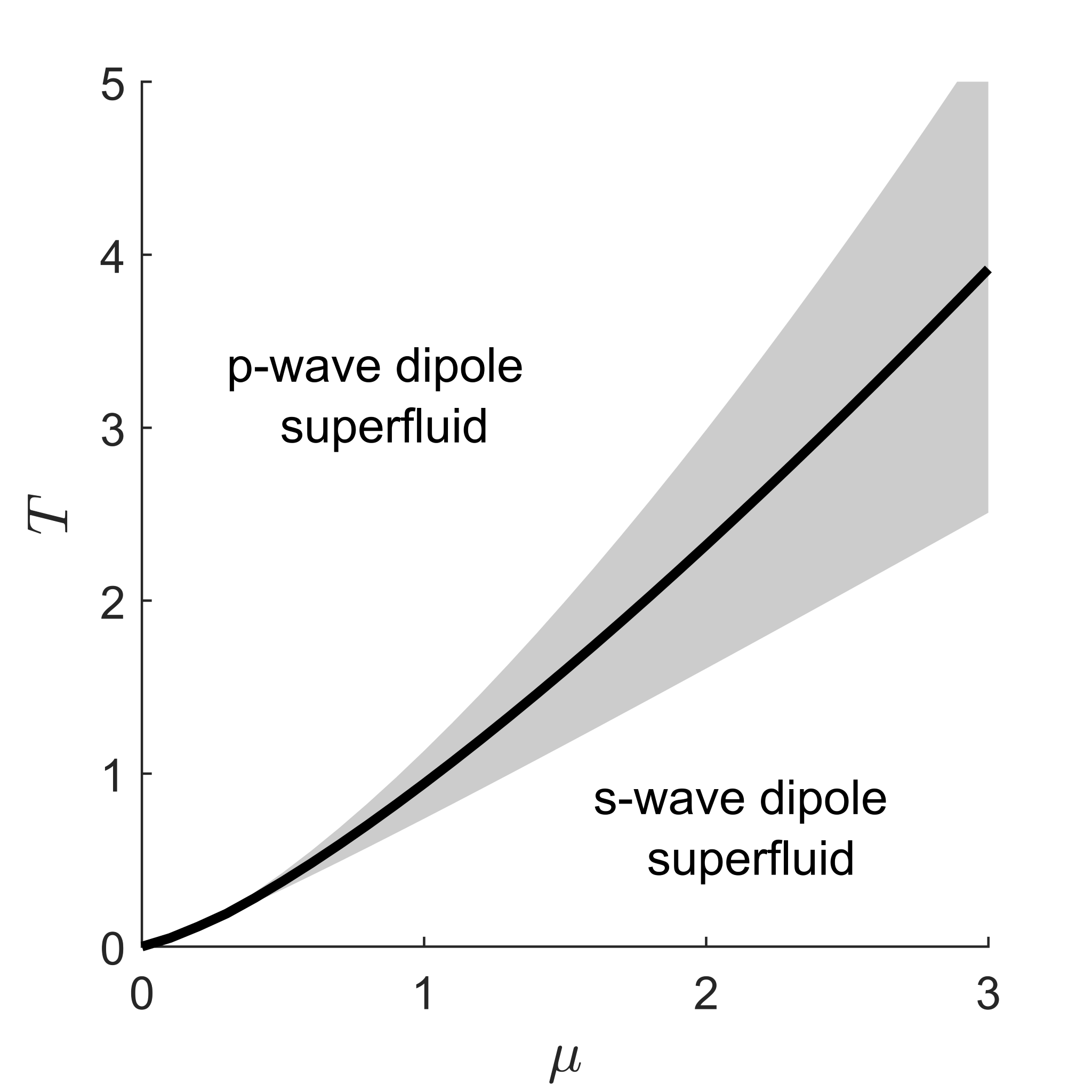}
    \caption{The phase diagram of model 1 in $d=3$ as a function of inverse temperature $\beta$ and chemical potential $\mu$ measured in natural units where $\text{Re}(\lambda) = \lambda_4 = 1$. The black line indicates a first order phase transition, and the shaded gray area is the region where both phases coexist.}
    \label{fig:mod1_pd}
\end{figure}

In Model 1 we find both phases in $d>2$ with a first-order transition between a high-temperature $p$-wave dipole superfluid and a low-temperature $s$-wave phase. In $d=2$ we only find the $p$-wave phase. See Fig.~\ref{fig:mod1_pt} for the thermal free energy at fixed chemical potential, and the phase diagram in Fig.~\ref{fig:mod1_pd}, both in $d=3$.

\begin{figure}
    \centering
    \includegraphics[width = 0.99\linewidth]{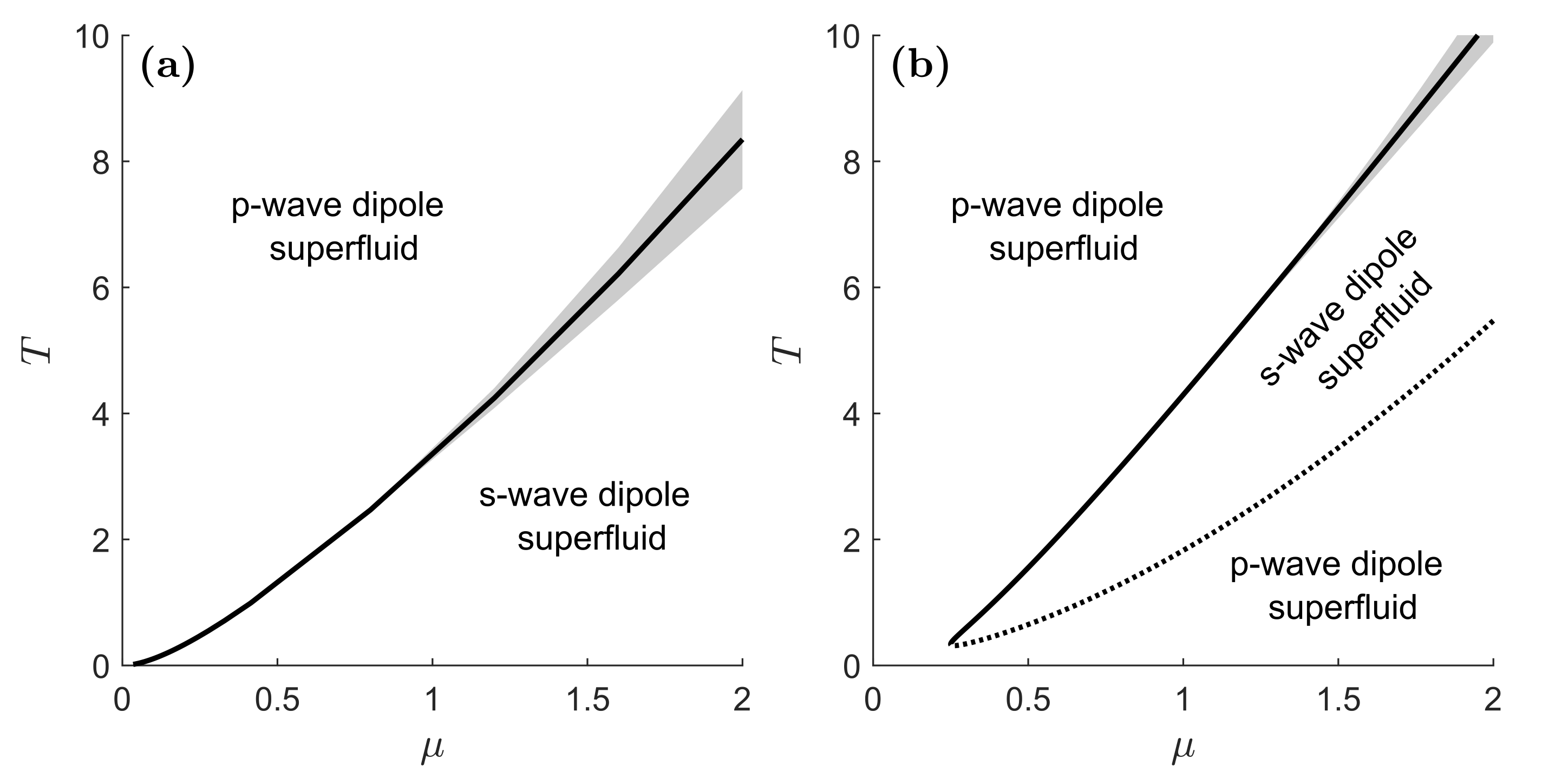}
    \caption{Sample phase diagrams of Model 2 in $d=3$ as a function of inverse temperature $\beta$ and chemical potential $\mu$ measured in natural units where $\lambda_S = 1$ and $\lambda_S|\gamma|^2+\lambda_4 = 1$. In Figure \textbf{(a)} $\text{Re}(\gamma) = 0.25$, while in Figure \textbf{(b)} $\text{Re}(\gamma) = -0.25$. The black line indicates a first order phase transition, the shaded gray area the region where both phases coexist, and the dotted black line a continuous phase transition.}
    \label{fig:mod2_pd}
\end{figure}

Though the same two phases exist in Model 2, the phase diagram for this model is more elaborate due to the additional parameters. When $\text{Re}(\gamma)>0$ the phase structure is similar to Model 1, with a first order transition from a high temperature $p$-wave phase to the $s$-wave phase at low temperature in $d>2$, and only the $p$-wave phase in $d=2$. However when $\text{Re}(\gamma)<0$ the low temperature phase is $p$-wave with $a_1 < 0$. This results in two possibilities, depending on the value of the chemical potential (assuming $\text{Re}(\gamma)$ is fixed.) If the chemical potential $\mu$ is small enough then there is no $s$-wave phase at all, and the system stays in the dipole superfluid phase at all temperatures. However if the chemical potential $\mu$ is increased beyond some critical threshold then there is an intermediate temperature range where the system is in the $s$-wave phase. Starting at high temperatures, the transition to this phase is first order, but the transition back to the dipole superfluid phase at low temperatures is continuous. Two sample phase diagrams for Model 2 are presented in Fig. \ref{fig:mod2_pd}, one for $\text{Re}(\gamma)>0$ and another for $\text{Re}(\gamma)<0$.

%--------------------------------------
\textit{Discussion.}~In this Letter we have solved continuum models of fractons with $N\gg1 $ degrees of freedom and a $U(N)$ global symmetry. These models have an exotic spacetime symmetry with a conserved dipole moment. We find two phases, a high-temperature phase in which dipole symmetry is spontaneously broken but the $U(N)$ is preserved, and a low-temperature phase in which $U(N)\to U(N-1)$ and dipole symmetry is also broken. The high-temperature phase has a vector order parameter, the gradient of the single-particle Green's function in the coincident limit, and the low-temperature phase has a scalar order parameter. So we dubbed the high-temperature phase a \emph{p-wave} dipole superfluid, and the low-temperature phase a \emph{s-wave} dipole superfluid.

We study many more features of these models in the Appendix, including the low-energy effective Goldstone descriptions in each phase and low-momentum response functions; soluble lattice versions of our models; the consequences of the dipole symmetry for the Hilbert space; subtleties associated with UV/IR mixing, and the quantization of dipole charges on a lattice; and solutions to the large $N$ Dyson equations in asymptotic limits. We refer the interested and intrepid reader there.

There are three lessons from our analysis that we highlight here, which we expect will be relevant for models with conserved dipole moment more generally.
\begin{enumerate}
	\item The quartic interactions involving spatial derivatives in these models are crucial and in general cannot be treated perturbatively. They generate a momentum-dependent self-energy, which tames loop integrals.
	\item The dipole symmetry is spontaneously broken. At high temperature, so that the ordinary global symmetry is unbroken, we find the low-energy Goldstone description of the continuum theory in the Appendix. This effective theory was proposed in~\cite{Lake:2022ico}, and is Gaussian in the deep IR. The low-temperature phase also has a Gaussian Goldstone description in the infrared, as proposed in~\cite{Pretko_2018}. These effective theories are perhaps best suited to describe systems with approximate dipole symmetry upon adding small breaking terms.
	\item The coarse-grained model is weakly sensitive to the details of an underlying lattice. We already encountered one such sensitivity. In finite volume, the dipole symmetry breaking implies the existence of a UV-sensitive ground state degeneracy. It arises through the volume of the zero modes associated with dipole breaking. That volume diverges in the continuum theory, but with a lattice regulator goes as the number of lattice sites. As we discuss in the Appendix, the lack of decoupling of the lattice can be studied in the Goldstone effective descriptions.
\end{enumerate}
We emphasize that we were able to study the large $N$ solution without difficulty in the continuum limit, and near the continuum limit.

We conclude with a brief list of future prospects. 

In this work we have identified the large $N$ solution of these models, including the large $N$ Green's function. It should be possible to compute the large $N$ four-point function, whose connected part comes from a sum over bubbles. From this one would be able to directly identify the low-energy Goldstone description, compute hydrodynamic response functions, find the free energy to $O(N^0)$, and more.

It has been argued in~\cite{Du:2021pbc} that the lowest Landau level may be thought of as a theory of fracton order with conserved dipole number, and conserved quadrupole trace. There are large $N$ models with this symmetry. A simple, but trivial example, is to consider a model with conserved dipole moment but vanishing trace of the dipole current, $J^i{}_i = 0$. In Model 2 this is achieved by setting $\lambda_S = 0$. Unfortunately in that case the large $N$ solution is trivial, with $\Sigma = -\mu$ at negative chemical potential and $\Sigma = 0$ at positive chemical potential. However, there are interesting large $N$ solutions in anisotropic models.

Relatedly, we may consider large $N$ fracton models with subsystem symmetry. We will report on that topic rather soon~\cite{Subsystem}.

\emph{ Acknowledgments.}
We would like to thank P.~Glorioso, A.~Gromov, A.~Karch, E.~Lake, and S.H.~Shao for enlightening discussions. KJ is supported in part by an NSERC Discovery Grant, and AR is supported by the U.S. Department of Energy under Grant No. DE-SC0022021 and a grant from the Simons Foundation (Grant 651678, AK).

\bibliographystyle{apsrev4-1}
\bibliography{refs}

\appendix

%--------------------------------------
\section{Dipole symmetry in the continuum}
%--------------------------------------
Let us consider models of $N$ complex scalar fields $\phi^a$ invariant under a global $U(N)$ symmetry. Let $Q$ be the generator of the $U(1)\subset U(N)$ center symmetry, which acts as $\phi^a \to e^{i\alpha} \phi^a$. We would like to study theories which also have a conserved dipole moment, invariant under the more general transformations $\phi^a \rightarrow e^{i\alpha + i d_i x^i} \phi^a$ for a dipole transformation $\vec{d}$. Notice that this dipole symmetry depends on the choice of origin, so the generators $D_i$ of dipole transformations do not commute with the momenta $P_j$. Indeed on $\phi$ we have the commutator
\begin{equation}
    \left[D_i,P_j \right] =i \delta_{ij} Q.
\end{equation}

Due to the dipole symmetry kinetic terms for the scalar with two spatial derivatives are forbidden. To see why this happens, it is instructive to move to momentum space where the dipole symmetry acts by $\tilde{\phi}(\omega,\vec{k}) \rightarrow \tilde{\phi}(\omega,\vec{k}+ \vec{d})$ for a dipole transformation $\vec{d}$. Note that $\vec{d}$ is valued in the space of momenta $\vec{k}$, which is non-compact in the continuum. The most general local and $U(N)$-invariant term in the action quadratic in $\phi$ with arbitrary derivatives takes the form $\int d\omega d^dk\, f(\omega,\vec{k})| \tilde{\phi}^a(\omega,\vec{k}) |^2$, which transforms as
\begin{align}
\begin{split}
    \int d\omega d^d k &\,f(\omega,\vec{k}) |\tilde{\phi}^a(\omega,\vec{k})|^2 
    \\
    &  \rightarrow \int d\omega d^d k \,f(\omega,\vec{k})|\tilde{\phi}^a(\omega,\vec{k}+\vec{d})|^2
     \\
    &\qquad = \int d\omega d^dk\, f(\omega,\vec{k}-\vec{d})| \tilde{\phi}^a(\omega,\vec{k})|^2, 
\end{split}
\end{align}
which is invariant if and only if $f(\omega,\vec{k}) = f(\omega)$ for all $\vec{k}$. This demonstrates the claim. 

As a result the simplest spatial kinetic term for a dipole conserving theory is at least quartic in the fields. In this Appendix we answer the question: what are the allowed terms with at most four powers of the field and four derivatives? 

Following \cite{Pretko_2018}, we can construct dipole-invariant field theories by introducing a ``covariant'' derivative $D_{ij}(\phi,\phi) = \partial_i \phi \partial_j \phi - \phi \partial_i \partial_j \phi$, which is covariant under the $U(1)$ and dipole symmetries. We note that we can also construct a generalization of this ``covariant'' derivative to include cases $D_{ij}(\phi_1,\phi_2)$ where $\phi_1$ and $\phi_2$ have different charges, as was done in \cite{jensen2021}:
\begin{equation} \label{E:D_gen}
    D_{ij}(\phi_1,\phi_2) = \frac{1}{2}\left(\frac{q_1}{q_2} \phi_1 \partial_i \partial_j \phi_2 - \partial_i\phi_1  \partial_j \phi_2  + (1 \leftrightarrow 2) \right),
\end{equation}
where $q_{i}$ is the $U(1)$ charge of $\phi_{i}$. This satisfies $D_{ij}(\phi_1,\phi_2) \to e^{i(q_1+q_2)(\alpha+d_ix^i)}D_{ij}(\phi_1,\phi_2)$. With $N$ scalars we may consider the objects $D_{ij}(\phi^a,\phi^b)$ and $D_{ij}(\bar{\phi}^a,\phi^b)$. However, when it comes to the latter we note that $D_{ij}(\phi_1,\phi_2) = - \frac{1}{2}\partial_i \partial_j (\phi_1\phi_2)$ when $q_1=-q_2$, so that $D_{ij}(\bar{\phi}^a,\phi^b) = -\frac{1}{2}\partial_i \partial_j(\bar{\phi}^a\phi^b)$.\footnote{Note that the dipole symmetric term $\int dt d^dx \,D_{ij}(\bar{\phi}^a,\phi^a)$ is a boundary term, consistent with the fact that there is no dipole-symmetric Lagrangian with two powers of the field and spatial derivatives. } Since the ordinary derivative of a charge singlet is also a charge/dipole singlet, it follows that when enumerating the list of dipole-invariant quantities, we need not consider $D_{ij}$ acting on objects of equal and opposite charge. That is we may form invariant quantities out of ordinary derivatives acting on singlets, the operator $D_{ij}$ acting on charged fields, and perhaps even more complicated operators generalizing $D_{ij}$ with more derivatives and fields.

If we impose invariance under rotations, $U(N)$, and dipole transformations, and neglect time derivatives, there are then six independent terms with four fields and up to four derivatives we can construct this way: the polynomial interaction $|\phi|^4 = (\bar{\phi}^a\phi^a)^2$, and
\begin{align}
	\nonumber
	&\text{Re}\left( \phi^a \phi^b D_{ii}\left(\bar{\phi}^a,\bar{\phi}^b\right)\right)\,, \qquad \text{Im}\left( \phi^a \phi^b D_{ii}(\bar{\phi}^a, \bar{\phi}^b)\right) \,,
	\\
	\nonumber
	&\partial_i\left( \bar{\phi}^a \phi^b \right) \partial_i \left( \bar{\phi}^b\phi^a\right)\,, \qquad \partial_i(\bar{\phi}^a\phi^a)\partial_i(\bar{\phi}^b\phi^b)\,,
	\\
	\nonumber
	& D_{ii}(\phi^a,\phi^b)D_{jj}(\bar{\phi}^a,\bar{\phi}^b) \,, \quad D_{\{ij\}}(\phi^a,\phi^b)D_{\{ij\}}(\bar{\phi}^b,\bar{\phi}^a)\,,
	\\
	\nonumber
	&\text{Re}\left(  D_{ii}(\phi^a\phi^b,D_{jj}(\bar{\phi}^a,\bar{\phi}^b))\right)\,,
	\\
	\label{E:dipole_terms}
	& \text{Im}\left( D_{ii}(\phi^a\phi^b,D_{jj}(\bar{\phi}^a,\bar{\phi}^b))\right)\,,
	\\
	\nonumber
	& \text{Re}\left( D_{\{ij\}}(\phi^a\phi^b,D_{\{ij\}}(\bar{\phi}^a,\bar{\phi}^b))\right)\,,
	\\
	\nonumber
	& \text{Im}\left( D_{\{ij\}}(\phi^a\phi^b,D_{\{ij\}}(\bar{\phi}^a\phi^b))\right)\,,
	\\
	\nonumber 
	& \nabla^2(\bar{\phi}^a\phi^b) \nabla^2(\bar{\phi}^b\phi^a)\,, \qquad\, \,\left(\nabla^2(\bar{\phi}^a\phi^a)\right)^2\,,
\end{align}
where $D_{\{ij\}}$ refers to the traceless part of $D_{ij}$. Four of these terms appeared in our Models 1 and 2 (the first, second, fifth, and sixth) although as we will discuss shortly, the large $N$ solutions of Models 1 and 2 also describe models in which the other terms are present

We would like to understand if these are the most general terms that preserve dipole symmetry, and if any of these terms are equivalent to the others under integration by parts. To address these questions it is simpler to work in momentum space, where the possible quartic terms can be fully classified by a momentum-dependent vertex. The most general quartic term invariant under $U(N)$ transformations and without time derivatives can be written as
\beq
	\int Dk_1Dk_2Dk_3 \,\bar{\tilde{\phi}}^a(k_1)\bar{\tilde{\phi}}^b(k_2)\tilde{\phi}^a(k_3)\tilde{\phi}^b(k_4)V_4(k_i) \,,
\eeq
where $V_4$ is the quartic vertex in momentum space and we shorthand $k_i = (\omega_i,\vec{k}_i)$ the frequency and momenta of the $i^{\rm th}$ field. Rotational invariance and parity imply that $V_4$ is only built from dot products, while the dipole symmetry implies it depends only on the differences in momenta $\vec{k}_{12} = \vec{k}_1-\vec{k}_2$ and $\vec{k}_{34}$ or the sums of momenta $\vec{k}_{1+3}=\vec{k}_1+\vec{k}_3, \vec{k}_{1+4}, \vec{k}_{2+3}$ and $\vec{k}_{2+4}$. It is symmetric under the simultaneous exchange $k_1\leftrightarrow k_2$ and $k_3\leftrightarrow k_4$, and reality imposes $V_4(k_1,k_2,k_3,k_4) = V_4^*(-k_3,-k_4,-k_1,-k_2)$. Lastly, momentum conservation implies $\sum_i k_i= 0$.

Momentum conservation allows to eliminate $\vec{k}_{1+3}, \vec{k}_{1+4}, \vec{k}_{2+3},$ and $\vec{k}_{2+4}$ in favor of the momentum differences $\vec{k}_{12}$ and $\vec{k}_{34}$. To find the most general quartic term invariant under the symmetries, we then find the most general function of $\vec{k}_{12}$ and $\vec{k}_{34}$ obeying the symmetry and reality conditions above. To fourth order in momenta, there are exactly nine solutions to those constraints (ignoring the constant):
\begin{align}
\begin{split}
	&|\vec{k}_{12}|^2 + |\vec{k}_{34}|^2\,, \quad i\left( |\vec{k}_{12}|^2 - |\vec{k}_{34}|^2\right)\,, \quad \vec{k}_{12}\cdot \vec{k}_{34}\,,
	\\
	& |\vec{k}_{12}|^2|\vec{k}_{34}|^2\,, \quad \,\,\,\,\,\,(\vec{k}_{12}\cdot \vec{k}_{34})^2\,,
	\\
	& |\vec{k}_{12}|^4 + |\vec{k}_{34}|^4\,, \quad i\left( |\vec{k}_{12}|^4 - |\vec{k}_{34}|^4\right)\,,
	\\
	& (\vec{k}_{12}\cdot \vec{k}_{34})(|\vec{k}_{12}|^2+|\vec{k}_{34}|^2)\,, 
	\\
	& i (\vec{k}_{12}\cdot \vec{k}_{34})(|\vec{k}_{12}|^2-|\vec{k}_{34}|^2)\,.
\end{split}
\end{align}
All of these momentum structures are encoded in the terms present in Eq.~\eqref{E:dipole_terms}. Note that Model 1, which has terms with at most two spatial derivatives, lacks the third two-derivative structure above, while Model 2, which has terms with at most four spatial derivatives, has the first second, fourth, and fifth.

However, recall that the Dyson equations involve only the forward scattering limit of the vertex $V_4$, for which $k_3 = -k_1=-k$ and $k_4 = -k_2=-k'$. In that limit the non-vanishing quartic vertices become
\beq
	|\vec{k}-\vec{k}'|^2\,, \qquad |\vec{k}-\vec{k}'|^4\,,
\eeq
which are already subsumed in our analysis. That is, had we included the additional quartic terms allowed by dipole symmetry in our Models, we would find precisely the same set of large $N$ solutions as without, just with a relabeling of coupling constants.

%--------------------------------------
\section{Details of the collective field formulation} \label{app:derivation}
%--------------------------------------
In this Appendix we give a detailed derivation of the collective  field formulation of Models 1 and 2, whose actions are given in \eqref{eq:action_model_1} and \eqref{eq:action_model_2}. In fact we consider a more general $U(N)$ vector model with arbitrary momentum-dependent quartic interaction with an action
\begin{align}
    S &= \int Dk\, \left(i\omega_n - \mu \right) \bar{\tilde{\phi}}^a(-k)\tilde{\phi}^a(k) 
    \\
    \nonumber
    &+\int Dk_1 Dk_2 Dk_3 \, \bar{\tilde{\phi}}^a(k_1) \bar{\tilde{\phi}}^b(k_2) \phi^a(k_3) \phi^a(k_4)V_4(k_i)\,,
\end{align}
where $\omega_n = 2\pi n/\beta$ are the Matsubara frequencies, the integration measure is
\begin{equation}
    \int Dk = \frac{1}{\beta}\sum_{n\in\mathbb{Z}}\int \frac{d^dk}{(2\pi)^d}\,,
\end{equation}
and $k_4 = -k_1-k_2-k_3$ by momentum conservation. The precise form of the quartic interaction $V_4(k_i)$ is given in the text, though it is not needed to derive the Dyson equations. 

We now perform a generalized Hubbard-Stratonovich transformation by integrating in a bilocal field $G(k,k') =  \bar{\phi}^a(k) \phi^a(k') $ and a Lagrange multiplier field $\Sigma(k,k')$ that enforces this definition, by inserting a resolution of the identity into the integral over the $\phi^a$,
\begin{align}
    1 &= \int DG D\Sigma
    \\
    \nonumber
    &  \exp\left[\int Dk Dk' \Sigma(k,k')\left( G(k,k') - \bar{\phi}^a(k) \phi^a(k') \right) \right] \,.
\end{align}
For a discussion of the details involved in the integration contour over $G$ and $\Sigma$ see~\cite{Aharony:2020omh} in the context of ordinary vector models. By integrating in $G$ we may rewrite the quartic term in $\phi$ as a quadratic one in terms of $G$, so that the full action is quadratic in $\phi$ and reads
\begin{align}
	S & = \int Dk \, (i\omega_n +\Sigma(-k,k))\bar{\tilde{\phi}}^a(-k)\tilde{\phi}^a(k) 
	\\
	\nonumber
	&- \mu \int Dk\,G(-k,k)   - \int Dk_1 Dk_2 \,\Sigma(k_1,k_2)G(k_1,k_2)
	\\
	\nonumber
	& + \int Dk_1Dk_2Dk_3\,G(k_1,k_3) G(k_2,k_4)V_4(k_i) \,.
\end{align}
Integrating out all but one of the scalar fields, $\phi^{a=1} \equiv \sigma$, we get the action
\begin{widetext} 
\begin{equation}
\begin{split}
    S' &= (N-1) \ln \det \left(i\omega_n\Delta +\Sigma \right)  -  \mu \int Dk\, G(-k,k) -   \int Dk_1 Dk_2 \,\Sigma(k_1,k_2) G(k_1,k_2)\\
    &+  \int Dk_1 Dk_2 Dk_3 \, G(k_1,k_3) G(k_2,k_4)V_4(k_i)
    + \int Dk_1 Dk_2 \left( i\omega_2 \Delta(k_1,k_2)+\Sigma(k_1,k_2)\right) \bar{\tilde{\sigma}}(-k) \tilde{\sigma}(k)\,,
\end{split}
\end{equation}
\end{widetext}
where $\Delta(k_1,k_2) \equiv \beta(2\pi)^{d} \delta_{n_1n_2}\delta^{(d)}(\vec{k}_1+\vec{k}_2)$ is a generalized identity operator (or $\delta$-function.)  

We now make a transitionally invariant ansatz the various fields, 
\begin{align}
\begin{split}
    \langle G(k,k') \rangle &= N\Delta(k,k')G(k')\,, 
    \\
     \langle\Sigma(k,k') \rangle &= \Delta(k,k')\Sigma(k')\,, 
\end{split}
\end{align}
and $\langle\sigma(x)\rangle = \sigma$ for some constant $\sigma$. The effective action evaluated on this ansatz reads
\begin{align}
    \frac{S'}{\beta V} = &(N-1)\int Dk \,\ln  \left(i\omega_n +\Sigma(k) \right) 
    \\
    \nonumber
    &+N \int Dk_1 Dk_2 \,G(k_1) G(k_2) NV_4(-k_1,-k_2,k_1,k_2)
    \\
    \nonumber
    &- N \int Dk\,(\Sigma(k)+\mu) G(k) + |\sigma|^2\Sigma(k=0)\,,
\end{align}
where we interpret $(2\pi)^d \delta^d(\vec{k}=\vec{0})$ as the spatial volume $V$. Note that at large $N$ with $(G(k),\Sigma(k)) = O(N^0)$ and $|\sigma|^2 =O(N)$, the action is $O(N)$, so that $(G,\Sigma;\sigma)$ are weakly coupled degrees of freedom at large $N$. Varying this effective action with respect to $G$ and $\Sigma$ gives the Dyson equations \eqref{E:dyson}, while the variation with respect to $\sigma$ results in the constraint $\sigma\Sigma(k=0) = 0$.

Next we would like to compute the on-shell action, the action evaluated on the solutions to the Dyson equations \eqref{E:dyson}. For those control parameters where only one large $N$ solution exists, the on-shell action computes $\beta F$ with $F$ the free energy. When two or more solutions exist, the solution with the lowest on-shell action dominates and again the action computes the free energy. 

Assuming ($G$, $\Sigma;\sigma)$ satisfy the Dyson equations, we can rewrite the the large $N$ on-shell action as
\begin{equation}
\begin{split}
    \frac{S'_{\text{on-shell}}}{N\beta V} = &\int Dk \,\ln  \left(i\omega_n +\Sigma(k) \right) 
    \\
    &-  \frac{1}{2}\int Dk(\Sigma(k)+\mu) G(k) \,.
\end{split}
\end{equation}
Next, using $G(k) = \frac{1}{i\omega_n+\Sigma(k)}$ and that $\Sigma$ only depends on spatial momentum, we perform the sums over Masubara modes. The two sums we need to evaluate are
\begin{equation}
       \frac{1}{\beta} \sum_{n\in \mathbb{Z}} \frac{1}{i\omega_n +\Sigma} = 
        \frac{1}{\beta} \sum_{n\in \mathbb{Z}} \frac{\Sigma}{\omega_n^2 +\Sigma^2}
        = \frac{1}{2} \coth\left( \frac{\beta \Sigma}{2}\right),
\end{equation}
and
\begin{equation}
    I = \frac{1}{\beta} \sum_{n\in \mathbb{Z}} \ln  \left(i\omega_n +\Sigma \right) = 
        \frac{1}{2\beta}\sum_{n\in \mathbb{Z}} \ln  \left(\omega_n^2 +\Sigma^2 \right),
\end{equation}
which is divergent. However this sum satisfies \cite{dolan1974}
\begin{equation}
    \frac{\partial I}{\partial \Sigma}  = 
        \frac{1}{\beta} \sum_{n\in \mathbb{Z}} \frac{\Sigma}{\omega_n^2 +\Sigma^2}
        = \frac{1}{2} \coth\left( \frac{\beta \Sigma}{2}\right),
\end{equation}
so 
\begin{equation}
    I = \frac{\Sigma}{2} + \frac{1}{\beta}\ln\left(1 -e^{-\beta \Sigma}\right) 
\end{equation}
up to some $\Sigma$-independent constant.

After performing the sum over Matsubara frequencies, we still need to regularize the momentum integrals appearing in the Dyson equations and the on-shell action. In particular we would like to regularize the momentum integrals in such a way as to preserve the dipole symmetry. For the rotationally-invariant solutions we can use dimensional regularization. By Veltman’s formula \cite{leibbrandt1975}
\begin{equation}
    \int d^dk\, k^z = 0\,.
\end{equation}
we have e.g. $\int d^d k \,\Sigma = 0 $, as $\Sigma$ is a polynomial in $k$. Similarly,
\begin{equation}
    \int d^dk\, (\Sigma+\mu)  \coth\left( \frac{\beta \Sigma}{2}\right) = \int d^dk (\Sigma+\mu)  \left[\coth\left( \frac{\beta \Sigma}{2}\right) - 1 \right] ,
\end{equation}
under dimensional regularization, where the integral on the right-hand side is finite. This same summation and regularization procedure allows us to write a regularized version of the Dyson equation~\eqref{E:Dyson2} for the self-energy,
\begin{align}
	\nonumber
	\Sigma(k) & =  \int \frac{d^dk'}{(2\pi)^d} NV_4(-k,-k',k,k') \left( \coth\left( \frac{\beta \Sigma(\vec{k}')}{2}\right)-1\right)
	\\
	\label{E:regularizedDyson}
	&\qquad  -\mu + 2\frac{|\sigma|^2}{N}NV_4(-k,0,k,0)\,.
\end{align}
as well a regularized version of the on-shell action
\begin{widetext}
\begin{equation}
\label{E:regularizedAction}
    \frac{S'_{\text{on-shell}}}{N V} = \int \frac{d^dk}{(2\pi)^d} \ln  \left(1 - e^{-\beta \Sigma} \right) 
    -  \frac{\beta}{4}\int \frac{d^dk}{(2\pi)^d} (\Sigma+\mu) \left[\coth\left(\frac{\beta \Sigma}{2} \right) - 1 \right]
    - \frac{\mu \beta}{2} \frac{|\sigma|^2}{N}\,.
\end{equation}
\end{widetext}

However, dimensional regularization is a bit unsatisfying, since by the dipole symmetry we can map a rotationally-invariant solution to a non-rotationally invariant one. A better choice of regulator is a version of Pauli-Villars regularization adapted to the quartic interactions of the model. Consider a model of scalars $\phi^a$ and heavy partner Grassmann-odd fields $\Psi^a$ with an action
\begin{align}
\begin{split}
	S =& \int Dk ( i\omega_n-\mu) \mathcal{G}(-k,k) + M \bar{\tilde{\Psi}}^a(-k)\tilde{\Psi}^a(k)
	\\
	&+ \int Dk_1 Dk_2Dk_3 \,\mathcal{G}(k_1,k_3)\mathcal{G}(k_2,k_4) V_4(k_i)\,,
\end{split}
\end{align}
where $\mathcal{G}(k_1,k_2) = \bar{\tilde{\phi}}^a(k_1)\tilde{\phi}^a(k_2) + \bar{\tilde{\Psi}}^a(k_1)\tilde{\Psi}^a(k_2)$. The parameter $M$ is a regulator mass which we will take to infinity at the end. We can then integrate in bilocal fields $(G,\Sigma)$ so as to set $G(k_1,k_2) = \mathcal{G}(k_1,k_2)$. Integrating out all scalars but $\phi^{a=1}=\sigma$ and all of the partner fields produces again a collective field theory of the fields $(G,\Sigma;\sigma)$, but which after making a translationally invariant ansatz becomes (approximating $N-1\approx N$)
\begin{align}
	\frac{S}{N\beta V}  =& \int Dk\ln \left( \frac{i\omega_n +\Sigma}{i\omega_n+\Sigma+M}\right)
	\\
	\nonumber
	& + \int Dk Dk' \,G(k)G(k')NV_4(-k,-k',k,k')
	\\
	\nonumber
	& - \int Dk\,(\Sigma(k)+\mu)G(k) + \frac{|\sigma|^2}{N}\Sigma(k=0)
\end{align}
The Dyson equations then take the form
\begin{align}
\begin{split}
	G(k) &= \frac{1}{i\omega_n+\Sigma(k)} - \frac{1}{i\omega_n+\Sigma(k)+M}
	\\
	& \qquad + \frac{|\sigma|^2}{N} \beta(2\pi)^d \delta_{n0}\delta^d(\vec{k})\,,
	\\
	\Sigma(k) & = -\mu + 2 \int Dk' \,G(k')NV_4(-k,-k',k,k')\,,
\end{split}
\end{align}
along with $\sigma \Sigma(k=0) = 0$. As before, these equations imply that the self-energy is exactly independent of frequency, so that we may perform the sum over Matsubara modes to produce a single Dyson equation for the self-energy,
\begin{align}
	\nonumber
	\Sigma(k) & = \int \frac{d^dk'}{(2\pi)^d} NV_4(-k,-k',k,k')
	\\
	\nonumber
	& \times  \left( \coth\left( \frac{\beta \Sigma(\vec{k}')}{2}\right)-\coth\left( \frac{\beta(\Sigma(\vec{k}')+M)}{2}\right)\right) 
	\\
	& \quad -\mu + 2\frac{|\sigma|^2}{N}NV_4(-k,0,k,0)\,.
\end{align}
Taking $M\to\infty$ gives the same Dyson equation~\eqref{E:regularizedDyson} we found via dimensional regularization for a rotationally-invariant configuration. The on-shell action for the Pauli-Villars regulated theory also produces the regularized action~\eqref{E:regularizedAction} in the $M\to\infty$ limit.

We can evaluate this on-shell action for the numerical solutions to the Dyson equations via numerical integration. However, we can also evaluate the integrals analytically for Model 1, resulting in 
\begin{equation}
\begin{split}
    \frac{S'_{\text{on-shell}}}{N V} &= 
    -\frac{ \beta (a_0+\mu)}{2 (4\pi \beta a_1)^{d/2} } \text{Li}_{\frac{d}{2}}\left(e^{-\beta a_0}\right) \\
    & -\frac{ d +4}{4 (4\pi \beta a_1)^{d/2}} 
     \text{Li}_{\frac{d+2}{2}}\left(e^{-\beta a_0}\right) 
    - \frac{\mu \beta}{2} \frac{|\sigma|^2}{N} .
\end{split}
\end{equation}

We wrap up this Appendix with two more brief observations concerning quantities we can compute in our Models from the large $N$ solutions to the Dyson equations. First, the regularized $U(1)$ charge density $\rho = \frac{1}{N}\langle \bar{\phi}^a\phi^a\rangle $ can be written as 
\begin{align}
\begin{split}
	\rho &= \int Dk \,G(k) 
	\\
	&= \frac{1}{2}\int \frac{d^dk}{(2\pi)^d} \left( \coth\left( \frac{\beta \Sigma(k)}{2}\right)-1\right) + \frac{|\sigma|^2}{N}\,.
\end{split}
\end{align}
In Model 1 we can use the Dyson equations to rewrite this as 
\beq
	\rho = 2 \text{Re}(\lambda)a_1\,,
\eeq
and in Model 2,
\beq
	\rho = \frac{ \lambda_S a_2}{2}\,.
\eeq
Note that the charge density receives two contributions: one is completely local, coming from the condensate, while the other comes from the non-local part of the large $N$ Green's function. This is reminiscent of Landau's two-fluid model for superfluidity, where there is a ``normal'' component of the charge density coming from the non-local part of the Green's function and a ``superfluid'' component from the condensate, and it suggests a two-fluid hydrodynamic description of the low-temperature phase. Later we showcase Goldstone effective descriptions of our Models, able to describe equilibrium finite-temperature states. However a full treatment of such a two-component ``fracton hydrodynamics'' is beyond the present work.

Lastly, consider the dipole current $J^{ij}$. By symmetry it may have a nonzero one-point function,
\beq
	\langle J^{ij}\rangle = N\rho_d \delta^{ij}\,.
\eeq
To compute the equilibrium dipole current $\rho_d$ we first couple the dipole currents of Models 1 and 2 to an external field $A_{ij}$ as in~\cite{Pretko_2018,jensen2021}. In practice this amounts to replacing the ``covariant derivative'' $D_{ij}(\phi_1,\phi_2)$ with
\begin{align}
	\nonumber
	D_{ij}(\phi_1,\phi_2)& = \frac{1}{2}\left( \frac{q_1}{q_2}\phi_1\partial_i \partial_j \phi_2 - \partial_i \phi_1\partial_j \phi_2 +(1\leftrightarrow 2)\right) 
	\\
	& \qquad + \frac{i}{4}(q_1+q_2)A_{ij}\phi_1\phi_2\,,
\end{align}
and the current is given by
\beq
	J^{ij} = 2 \frac{\delta S}{\delta A_{ij}}\,.
\eeq
In Model 1, deforming by the coupling $A_{ij}$ modifies the coefficient $\lambda_4$ of the quartic term,
\beq
\label{E:dipoleDeformation}
	S_{\rm Model 1} \to S_{\rm Model 1} + \int d^{d+1}x \,\frac{\text{Im}(\lambda)}{N}(\bar{\phi}^a\phi^a)^2A_{ii}\,,
\eeq
so that
\beq
	\lambda_4 \to \lambda_4 + \text{Im}(\lambda)A_{ii}\,.
\eeq
So $\rho_d$ can be computed by a derivative of the on-shell action with respect to $\lambda_4$. Alternatively, from~\eqref{E:dipoleDeformation}, we have
\beq
	J^{ij} =2 \frac{\text{Im}(\lambda)}{N}(\bar{\phi}^a\phi^a)^2\delta^{ij}\,,
\eeq
so that
\begin{align}
	\nonumber
	\rho_d &= 2\text{Im}(\lambda) \left\langle \left(\frac{1}{N} \bar{\phi}^a\phi^a\right)\left(\frac{1}{N}\bar{\phi}^b\phi^b\right)\right\rangle
	\\
	\nonumber
	& = 2\text{Im}(\lambda)\left( \left\langle \frac{1}{N}\bar{\phi}^a\phi^a\right\rangle\left\langle \frac{1}{N}\bar{\phi}^b\phi^b\right\rangle+O\left( \frac{1}{N}\right)\right)
	\\
	& =2 \text{Im}(\lambda)\rho^2 + O\left( \frac{1}{N}\right)\,,
\end{align}
where we have exploited large $N$ factorization. Note that the coupling $\text{Im}(\lambda)$, which drops out of the large $N$ Dyson equations, controls the strength of the equilibrium dipole current. 

Similar statements hold in Model 2.

%--------------------------------------
\section{Some more details about Model 2}
%--------------------------------------
Here we give a more complete analysis of Model $2$, starting with the Dyson equations. While we gave the full Dyson equations for Model 1 in the text, the Dyson equations for Model 2 are more complicated as the integrals cannot be evaluated analytically, and are summarized by
\begin{widetext} 
\begin{equation} \label{eq:SD_mod2}
    \begin{aligned}
     a_0 &= -\mu +  a_2 \left(|\gamma|^2 + \frac{\lambda_4}{\lambda_S} \right)+ \frac{d~\text{Re}(\gamma)}{d+2} (a_1 - 2 \text{Re}(\gamma) a_2 )
     + A_d \lambda_S \int_0^\infty dk \,{k}^{d+3}
     \left( \coth\left( \frac{\beta(a_0 + a_1{k}^2 + a_2 {k}^4 )}{2}\right)-1\right),  \\
     a_1 &= 2 \text{Re}(\gamma) a_2 + A_d \lambda_S \frac{2(d+2)}{d} \int_0^\infty dk \,{k}^{d+1} \left( \coth\left( \frac{\beta(a_0 + a_1{k}^2 + a_2 {k}^4)}{2}\right)-1\right), \\
     a_2 &= 2 \lambda_S\frac{|\sigma|^2}{N} + A_d \lambda_S\int_0^\infty dk \,{k}^{d-1}      \left( \coth\left( \frac{\beta(a_0 + a_1{k}^2 + a_2 {k}^4 )}{2}\right)-1\right),
    \end{aligned} 
\end{equation}
\end{widetext}
where $A_d =  \frac{2}{(4\pi)^{d/2} \Gamma(d/2)}$ is the normalized volume of the $d-1$ sphere, and the additional condition $\sigma a_0 = 0$.

\begin{figure*}
    \centering
    \includegraphics[width = 0.97\linewidth]{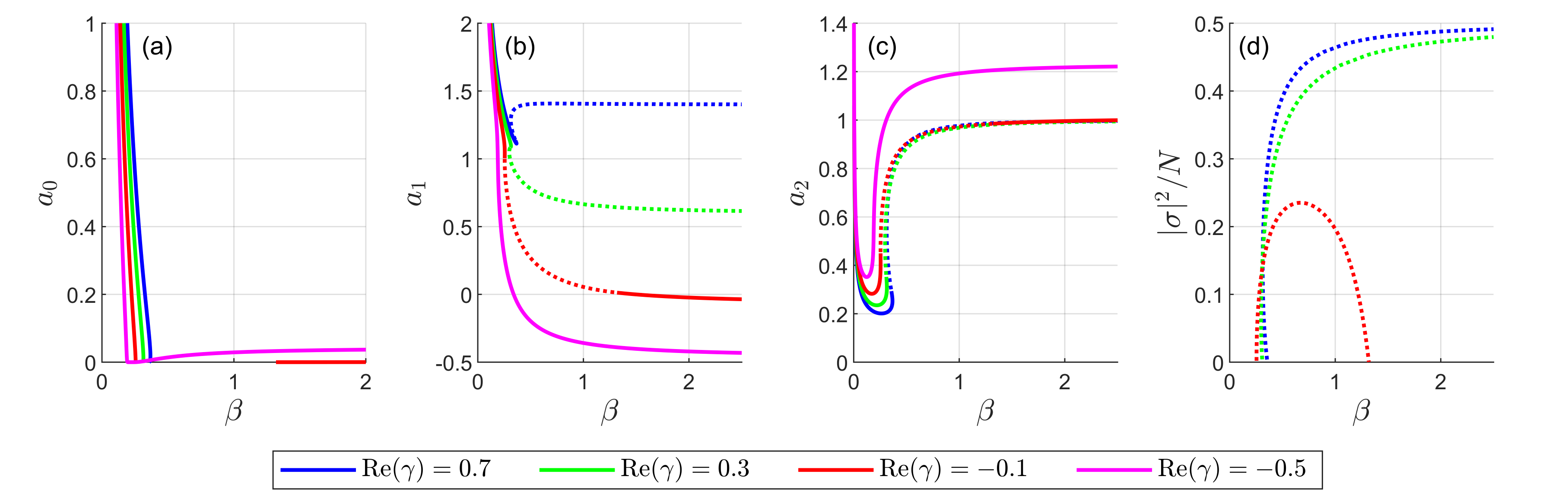}
    \caption{Solutions to the Dyson equations for Model 2, \eqref{eq:SD_mod2}, as a function of inverse temperature $\beta$  in $d=3$, with $\mu = 1$, and for several values of $\text{Re}(\gamma)$. The solid lines represent the $p$-wave phase and the dashed lines the $s$-wave phase. In all cases we are working in natural units where $\lambda_S = |\gamma|^2 +\lambda_4/\lambda_S = 1$.}
    \label{fig:mod2_pt}
\end{figure*}

\begin{figure}
    \centering
    \includegraphics[width = 0.97\linewidth]{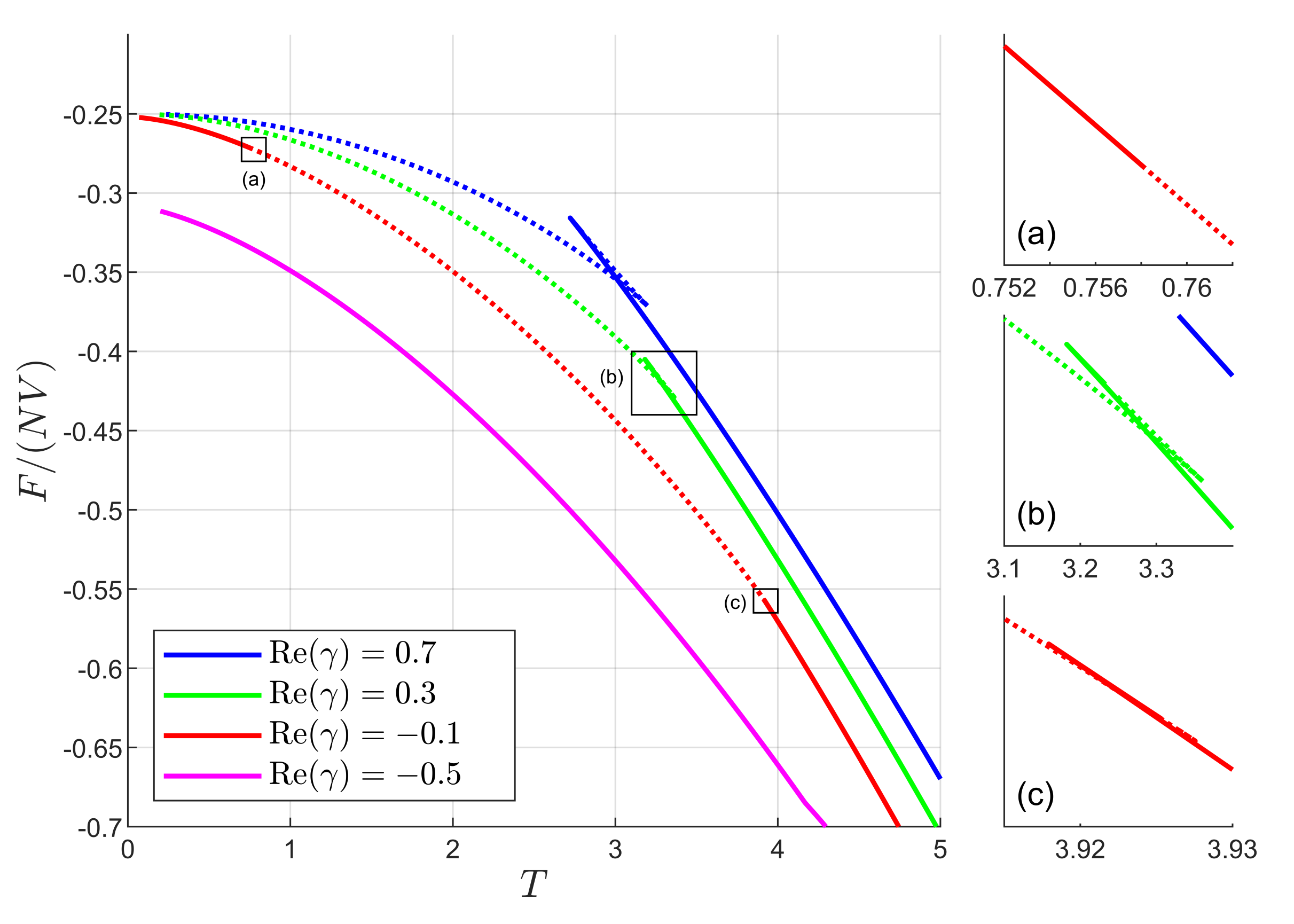}
    \caption{The large $N$ free energy density for model 2, \eqref{E:regularizedAction}, as a function of temperature $T$  in $d=3$, with $\mu = 1$, and for several values of $\text{Re}(\gamma)$. The solid lines represent the $p$-wave phase and the dashed lines the $s$-wave phase. In all cases we are working in natural units where $\lambda_S = |\gamma|^2 +\lambda_4/\lambda_S = 1$. The side panels are zoomed in on the phase transition.}
    \label{fig:mod2_fe}
\end{figure}

While the integrals in \eqref{eq:SD_mod2} cannot be evaluated analytically, these three coupled equations can easily be solved numerically for $(a_0,a_1,a_2;\sigma)$ for given values of the parameters. Though the Dyson equations seem a priori to depend on the parameters and couplings independently, we can nondimensionalize these equations to see that there are only three independent parameters. We choose the independent parameters to be $\mu, \beta$ and $\text{Re}(\gamma)$, while working in units where $\lambda_S = |\gamma|^2+\lambda_4/\lambda_S = 1$. It is fairly straightforward to see that $\lambda_4$, $a_i$ and $\mu$ scale like $\lambda_S^1$, while $\beta$ scales like $\lambda_S^{-1}$, and the rest of the parameters do not scale with $\lambda_S$. The re-scaling of $\alpha \equiv |\gamma|^2 +\lambda_4/\lambda_S$ is more involved, and comes from the freedom to re-scale the integration parameter $k$. In general one can define the dimensionless quantities $\tilde{a}_i, \tilde{\mu},\tilde{\beta},\tilde{\sigma},$ and $\text{Re}(\tilde{\gamma})$ via the relations
\begin{align}
\nonumber
    &a_j = \lambda_S \alpha^{d/4 + 1 - j/2}\tilde{a}_j\,,
    &&\mu = \lambda_S \alpha^{d/4 + 1}\tilde{\mu}\,,
     \\
    &\beta = \lambda_S^{-1} \alpha^{-d/4 - 1}\tilde{\beta}\,,
    &&\text{Re}(\gamma) = \alpha^{1/2}\text{Re}(\tilde{\gamma})\,, 
    \\
    \nonumber
    &|\sigma|^2 = \alpha^{d/4}|\tilde{\sigma}|^2\,,
\end{align}
and rescale the integration variable $k = \alpha^{1/4} \tilde{k}$, to get a dimensionless version of the Dyson equations, which is the same as working in natural units where $\lambda_S = 1$ and $\alpha \equiv |\gamma|^2 +\lambda_4/\lambda_S = 1$.

We present some numerical solutions to the Dyson equations of Model 2 in Fig. \ref{fig:mod2_pt}. We see that these solutions are qualitatively different than those of Model 1, especially when $\text{Re}(\gamma)<0$. In particular, when $\text{Re}(\gamma)<0$ the low temperature phase is a $p$-wave dipole superfluid, with a possible $p$-wave phase at intermediate temperatures. This corresponds to a much richer phase diagram which was presented in Fig. \ref{fig:mod2_pd}.

As in Model 1, the phase transition from the high temperature $p$-wave phase to the low temperature $s$-wave phase is a first order transition. However the second phase transition is continuous (when it exists.) The numerics seem to imply that free energy around the transition has a smooth second derivative, and so this transition is higher than second order. The noise introduced from numerical differentiation makes higher derivatives too noisy, so it hard to evaluate numerically the order of this phase transition.

%--------------------------------------
\section{Dipole symmetry on the lattice}
%--------------------------------------
In this Appendix we consider various aspects of lattice models with a conserved dipole moment. See~\cite{Gorantla:2022eem} for a thorough discussion of one-dimensional models with these symmetries. Here we have in mind quantum mechanical models on a spatial lattice $\Gamma$ generated by a basis of lattice vectors $\hat{a}_{m=1,2,\hdots,d}$, where ordinary charge $Q$ is also quantized and there is a symmetry under discrete lattice translations.

In this setting dipole moments are quantized, since the dipole moment is just a sum of charges times the lattice vectors where those charges are placed. Picking units so that the $U(1)$ charges are integers, dipole moments $\vec{D}$ are valued in $\Gamma \mathbb{Z}$, with $\vec{D} = \sum_m \mathfrak{d}_m \hat{a}_m$ where the $\mathfrak{d}_m$ are integers. Dipole symmetry transformations are then valued in the dual lattice $\widetilde{\Gamma}$. This can be seen from the dipole transformation itself. For example, consider a model of a charged field $\phi(t,\vec{x})$ where $\vec{x} = \sum_m x_m \hat{a}_m$ is a lattice vector. Dipole transformations $\vec{d}$ act on $\phi$ as
\beq
	\phi(t,\vec{x}) \to e^{i \vec{d}\cdot \vec{x}} \phi(t,\vec{x})\,,
\eeq
so that $\vec{d}$ is valued in the space of momenta, $\widetilde{\Gamma}$. When the lattice is large but finite, so is the dipole group parameterized by $\vec{d}$, while on an infinite lattice, $\widetilde{\Gamma}$ is continuous.

Conversely, the dipole moments and dipole transformations are not quantized if either the ordinary charges are not quantized, and/or the model is formulated in the continuum.

We would like to understand the consequences of the dipole symmetry for the Hilbert space. Let $\mathcal{T}_m$ be the translation operator by a basis lattice vector $\hat{a}_m$. The symmetries of the problem imply that the $\mathcal{T}_m$, $Q$, and the dipole moment operator $\vec{D}$ all commute with the Hamiltonian. The lattice version of the dipole algebra includes the statement that
\beq
	\mathcal{T}_m^{\dagger} \vec{D} \mathcal{T}_m = \vec{D} + Q \hat{a}_m\,,
\eeq
where $\vec{D}$ refers to the dipole moment operator and $Q$ is the $U(1)$ charge operator. This is a very formal way of phrasing a simple physical fact: if we put a charge $Q$ at some lattice site, translating by $\hat{a}_m$ changes the dipole moment by $Q\hat{a}_m$. In the quantum theory, because lattice translations and dipole moment do not commute, we cannot simultaneously diagonalize all symmetry generators. Choosing to diagonalize the Hamiltonian, $Q$, and $\vec{D}$, the Hilbert space of such a model is spanned by kets of the form
\beq
	|E,Q,\vec{D}\rangle\,,
\eeq
where now $Q$ and $\vec{D}$ represent the eigenvalues of the state under $U(1)$ charge and dipole moment. The representation theory of the dipole symmetry depends sensitively on the total charge $Q$ of the state. Acting with a lattice translation on such a state shifts the dipole moment
\beq
	\mathcal{T}_m |E,Q,\vec{D}\rangle = |E,Q,\vec{D}+Q \hat{a}_m\rangle\,,
\eeq
while leaving the energy and $U(1)$ charge alone. Acting with translations on a neutral state generates no new states, but acting on a state of nonzero charge generates a large number of degenerate states with dipole moments $\vec{D}+Q \Gamma$. This leads to a representation of the dipole symmetry with dimension
\beq
	\text{dim} = \begin{cases} 1 \,, & Q = 0\,, \\ N_{\rm sites}\,, & Q \neq 0\,,\end{cases}
\eeq
where $N_{\rm sites}$ is the number of lattice sites.

Alternatively, we could diagonalize the Hamiltonian, $U(1)$ charge, and lattice momenta $\vec{P}$ coming from $\mathcal{T}_m = \exp\left(  -i \hat{a}_m \cdot \vec{P}\right)$, producing kets of the form
\beq
	|E,Q,\vec{P}\rangle\,.
\eeq
Dipole transformations $\vec{d}$ act on these states as
\beq
\label{E:dipoleActingOnMomentum}
	e^{i \vec{d}\cdot \vec{D}} |E,Q,\vec{P}\rangle = |E,Q,\vec{P}+Q \vec{d}\rangle\,,
\eeq
generating a large number of degenerate states with momenta $\vec{P}+Q\vec{d}$. Notice that the requirement that momenta are valued in the dual lattice constrains $\vec{d}$ to be valued in $\widetilde{\Gamma}$ too. This again leads to a representation of the dipole symmetry, a reorganization of the one above, with dimension equal to $1$ for $Q=0$ and the number of lattice sites for $Q\neq 0$. 

Said another way, if we consider a superselection sector of nonzero charge, there is one choice of basis where momentum is diagonalized and dipole symmetry is spontaneously broken. On the other hand, we could diagonalize dipole number, and then momentum is spontaneously broken. Either way leads to a large number $\propto N_{\rm sites}$ of ground states in that sector. This large degeneracy persists above the ground states and in general the density of states at fixed charge behaves as $\rho_Q(E)\sim N_{\rm sites}$. 

This feature has an important consequence for the partition function. In the canonical ensemble at fixed charge $Q\neq 0$ we trivially have 
\beq
	Z_Q = \text{tr}_{\mathcal{H}_Q} \left( e^{-\beta H}\right) = N_{\rm sites} \widetilde{Z}\,,
\eeq
since all of the states that contribute to $Z_Q$ fall into representations of the dipole symmetry with dimension $N_{\rm sites}$. 

These statements nicely match our findings at large $N$. Our large $N$ solutions are translationally invariant, and therefore diagonalize momentum. Dipole symmetry was spontaneously broken, and the partition function is proportional to the zero mode volume. That volume was the dimensionless volume of single-particle momenta, which was infinite in the continuum, but $\sim V V_{\rm VZ}\sim N_{\rm sites}$ in a very dense lattice regularization. Strictly speaking, our continuum models are unable to detect this fact, but in the next Appendix we write down soluble lattice versions of our large $N$ models which exhibit dipole symmetry breaking with this zero mode volume.

%--------------------------------------
\section{Soluble lattice models} 
%--------------------------------------
In this Appendix we consider soluble large $N$ Models on a cubic lattice. Let us consider a lattice version of Model 1 for simplicity. We have an $N$-component scalar field $\phi^a=\phi^a(\tau,\vec{x})$ where $\vec{x} = a(\ell_1 \hat{e}_1 + \ell_2 \hat{e}_2 + \hdots)$, $\hat{e}_i$ is a unit vector in the $i^{\rm th}$ direction, and the $\ell_i$ are integers. Dipole transformations act as
\beq
	\phi^a(\tau,\vec{x}) \to e^{i \vec{d}\cdot \vec{x}} \phi^a(\tau,\vec{x})\,,
\eeq
and we wish to write down models invariant under this transformation, constant $U(N)$ rotations, time translations, and lattice translations. 

Let us cut to the chase and write down the action we have in mind after integrating in collective fields $(G,\Sigma)$ as in Appendix B. It takes the same form as in the continuum,
\begin{align}
	S = &\int Dk \,\left(i\omega_n+\Sigma(-k,k)\right)\bar{\tilde{\phi}}^a(-k)\tilde{\phi}^a(k)
	\\
	\nonumber
	& - \mu \int Dk \,G(-k,k)- \int Dk_1 Dk_2 \,\Sigma(k_1,k_2)G(k_1,k_2)
	\\
	\nonumber
	& + \int Dk_1 Dk_2 Dk_3 \,G(k_1,k_3)G(k_2,k_4) V_4(k_i)\,,
\end{align}
except now
\beq
	NV_4(k_i) = \frac{2}{a^2}\sum_{i=1}^d\left(\lambda \sin^2\left( \frac{k_{12i}a}{2}\right)+\bar{\lambda}\sin^2\left(\frac{ k_{34i}a}{2}\right)\right) + \lambda_4\,,
\eeq
and the integrals over spatial momenta are understood to be properly normalized sums over lattice momenta. In this form the dipole symmetry is the invariance under
\begin{align}
\begin{split}
	&\vec{k}_{1,2} \to \vec{k}_{1,2} - \vec{d}\,,
	\\
	& \vec{k}_{3,4}\to \vec{k}_{3,4} + \vec{d}\,,
\end{split}
\end{align}
for $\vec{d}$ a dipole transformation. Note that $\vec{k}_{1,2}$ are the momenta of the anti-charges $\bar{\phi}$, and $\vec{k}_{3,4}$ the momenta of the charges $\phi$, so that this transformation law is exactly commensurate with the action~\eqref{E:dipoleActingOnMomentum}. In the continuum limit $a\to 0$ the quartic interaction becomes $NV_4=\frac{1}{2} (\lambda |\vec{k}_{12}|^2+\bar{\lambda}|\vec{k}_{34}|^2)+\lambda_4$, the interaction of the continuum model we discussed in the main text.

This Model is just as soluble as Model 1. In particular it is weakly coupled at large $N$ and the Dyson equations can be solved without much effort. Integrating out all scalars but the first $\phi^{a=1}=\sigma$ and making a translationally-invariant ansatz
\begin{align}
\begin{split}
	G(k_1,k_2) & = N \Delta_{k_1,k_2} G(k_2)\,,
	\\
	\Sigma(k_1,k_2) & = \Delta_{k_1,k_2}\Sigma(k_2)\,,
\end{split}
\end{align}
with $\Delta_{k_1,k_2} = \frac{1}{\beta} (2\pi)^d \delta_{n_1+n_2,0}\delta_{\vec{k}_1+\vec{k}_2,0}$ and $\sigma(\tau,\vec{x})=\sigma$, we arrive at the same Dyson equations as before, just with a new quartic vertex,
\begin{align}
\begin{split}
	G(k) & = \frac{1}{i\omega_n+\Sigma(k)} + \frac{|\sigma|^2}{N}\Delta_{k,0}\,,
	\\
	\Sigma(k) & = -\mu + 2\int Dk' \,G(k') NV_4(-k,-k',k,k')\,,
\end{split}
\end{align}
along with $\sigma \Sigma(k=0) = 0$. Since the quartic interaction is independent of frequency, we can sum over Matsubara modes to produce a Dyson equation for the self-energy alone,
\begin{align}
	\nonumber
	\Sigma(k) & = \int \frac{d^dk'}{(2\pi)^d}\, NV_4(-k,-k',k,k') \coth\left( \frac{\beta \Sigma(\vec{k}')}{2}\right)
	\\
	& - \mu + 2 \frac{|\sigma|^2}{N}NV_4(-k,0,k,0)\,,
\end{align}
which only differs from the continuum by the sum over momenta and the form of $V_4$.
To make contact with our continuum models we also include heavy regulator fields, so that $\coth\left( \frac{\beta \Sigma}{2}\right) \to \coth\left( \frac{\beta \Sigma}{2}\right)-1$.

The quartic interaction in Model 1 is a quadratic polynomial in the $e^{ik_ia}$. Thus $\Sigma$ is also a quadratic polynomial in the same. Making an ansatz that $\Sigma$ is invariant under the discrete $90^{\circ}$ rotations preserved by the lattice along with parity, we can parameterize $\Sigma$ by two constants $a_1$ and $a_0$ with
\beq
	\Sigma(k) = \frac{4a_1}{a^2}\sum_{i=1}^d \sin^2\left( \frac{k_i a}{2}\right) + a_0\,,
\eeq
so that the Dyson equation above becomes the coupled equations
\begin{align}
\begin{split}
	a_1 & =\text{Re}(\lambda)\int \frac{d^dk}{(2\pi)^d}\left(1 - \frac{2}{d}\sum_{i=1}^d \sin^2\left(\frac{k'_i a}{2}\right)\right) \varsigma
	\\
	& \qquad \qquad \qquad \qquad \qquad +  \frac{2|\sigma|^2 \text{Re}(\lambda)}{N}\,,
	\\
	a_0 & = \int \frac{d^dk}{(2\pi)^d} \left( \frac{4\text{Re}(\lambda)}{a^2}\sum_{i=1}^d \sin^2\left(\frac{k_ia}{2}\right)+\lambda_4\right)\varsigma
	\\
	& \qquad\qquad \qquad \qquad \qquad  - \mu + \frac{2|\sigma|^2\lambda_4}{N}\,,
\end{split}
\end{align}
where
\beq
	\varsigma = \coth\left( \frac{\beta \Sigma(\vec{k}')}{2}\right)-1\,,
\eeq
for $a_1$ and $a_0$, along with $\sigma a_0=0$. These equations are valid at finite lattice spacing, but are especially tractable in the $a\to 0$ limit so that the sums over lattice momenta are very well-approximated by integrals. In that regime we find a solution with $O(a^0)$ values of the parameters $(a_1,a_0;\sigma)$ of our ansatz, in which case these sums are dominated by the $|ka|\ll 1$ regime, and can be performed analytically. They then pass over to those~\eqref{eq:dyson_mod1_N} of the continuum model that we referenced in the main text. That is, we find solutions to the Dyson equations of the lattice model that have a continuum limit given by the large $N$ solution described in the main text.

However, that is not the whole story. In addition to that solution we find another one that is very sensitive to the lattice. It has $a_1 = 0$ exactly along with 
\beq
	 \beta a_0 \approx -(d+2)\ln a\,,
\eeq
in the limit of small lattice spacing. The ensuing Green's function is
\beq
	G \approx \frac{1}{i\omega_n -\frac{d+2}{\beta}\ln a}\,,
\eeq
which describes an insulating phase with a UV-sensitive gap $\sim -\frac{(d+2)}{\beta}\ln a$. However, the free energy density of this solution is $O(a^2\ln^2 a)$, so that this solution is subdominant compared to the solution we described in the main text for order one temperatures and chemical potentials. (Compare such a free energy with that presented in the right-most panel of Fig.~\ref{fig:mod1_pt}.) 

In conclusion, the solution of the lattice model has a continuum limit, namely the large $N$ solution described in the main text. We henceforth focus on that solution.

The dipole symmetry acts on that large $N$ Green's function and self-energy as $G(\omega_n,\vec{k}) \to G(\omega_n,\vec{k}+\vec{d})$ and $\Sigma(\vec{k}) \to \Sigma(\vec{k}+\vec{d})$ generating a family of solutions labeled by allowed dipole transformations $\vec{d}$. For a model on a lattice $\Gamma$, $\vec{d}$ is valued in the dual lattice $\widetilde{\Gamma}$. The number of transformations, and so large $N$ solutions related by dipole symmetry, is then equal to the number of lattice sites $N_{\rm sites} \sim V V_{\rm BZ}$.

From the point of view of the functional integral over $G$ and $\Sigma$, the dipole symmetry implies the existence of zero modes in the spectrum of fluctuations around a given solution $(G,\Sigma;\sigma)$. However for a finite lattice the would-be integral over these zero modes really produces a finite sum over dipole transformations. We can interpret this sum as the statement that we sum over $N_{\rm sites}$ different large $N$ solutions for $(G,\Sigma;\sigma)$ related by the dipole symmetry, and of course we integrate over continuous (non-zero mode) fluctuations around each saddle. This leads to a large $N$ partition function $Z \approx  N_{\rm sites} e^{-S}$ where $S$ is the action of the dominant large $N$ solution. This behavior is completely consistent with the discussion of the last Appendix, that states in the Hilbert space with nonzero charge fall into representations with dimension $N_{\rm sites}$.

We wrap up this Appendix with a discussion of operators charged under the dipole symmetry. In the continuum there are multi-local operators that carry dipole moment, the most basic of which is $G(x_1,x_2) = \bar{\phi}^a(x_1)\phi^a(x_2)$ which carries dipole moment $-\vec{x}_{12}$. However there are no local operators that carry dipole moment. One may consider objects like $\bar{\phi}^a(x_1) \partial_i \phi^a(x_2)$ in the coincident limit, but these do not rotate by a phase under dipole transformations.

Lattice models are another story. On the lattice we may consider bilocals like $G(x_1,x_2)$ where the insertions are a finite number of lattice spacings away from each other. Operators like this are ``local'' as $a\to 0$. By adjusting the separation $\vec{x}_{12}$ we can find $U(N)$-neutral operators that carry any amount of dipole charge $\vec{D}$ provided that $\vec{D}/a$ is finite as $a\to 0$. 

%--------------------------------------
\section{Nonzero velocity}
%--------------------------------------
In boost-invariant theories equilibrium states exist at nonzero velocity, but by a boost such states are equivalent to thermal states at rest. In non-boost-invariant but translation-invariant theories equilibrium states still generally exist at nonzero velocity, but they are unrelated by symmetry to states at rest. These states are characterized by a partition function with a chemical potential for momentum,
\beq
	Z=\text{tr}\left( e^{-\beta (H-\mu Q-\vec{u}\cdot \vec{P})}\right)\,,
\eeq
producing an equilibrium state at velocity $\vec{u}$. 

However, models of fracton order are not generic. There is an eternal conflict between velocity and charge in models with fracton order that have a conserved dipole moment. On physical grounds, a translation-invariant equilibrium state with a uniform charge density cannot move without changing the dipole moment. So we may consider an equilibrium state at nonzero velocity and zero charge, or zero velocity and nonzero charge, but not both nonzero. 

The partition function knows about this fact in the following way. Let us label states in the Hilbert space by simultaneously diagonalizing $(H,Q,\vec{D})$, producing kets of the form $|E,Q,\vec{D}\rangle$. The subspace of fixed energy and charge $\mathcal{H}_{E,Q}$ contributes to the partition function as
\beq
	\rho_{Q}(E) e^{-\beta (E-\mu Q)} \text{tr}_{\mathcal{H}_{E,Q}} \left( e^{\beta \vec{u}\cdot \vec{P}}\right) \,,
\eeq
but the matrix elements that contribute to this trace are
\beq
	\langle E,Q,\vec{D}|e^{\beta \vec{u}\cdot \vec{P}}|E,Q,\vec{D}\rangle = \langle E,Q,\vec{D}|E,Q,\vec{D}+ i Q\beta \vec{u}\rangle\,,
\eeq
which vanishes when $Q\vec{u} \neq \vec{0}$. So when $\vec{u} \neq \vec{0}$ the partition function localizes to a sum over states with $Q=0$. On the other hand, if $\vec{u}=\vec{0}$, states of all charges contribute.

This blood feud has enormous consequences for a theory of transport in these models, in particular it necessitates additional fields in the low-energy hydrodynamic description, which will be investigated in detail in~\cite{Hydro}. For now, we will show how this conflict between charge and velocity is realized in our large $N$ models.

Consider our large $N$ models in the imaginary time formalism at nonzero imaginary velocity $\vec{u}_E$. These theories then live on a Euclidean cylinder in which the identifications read
\beq
	(\tau,\vec{x})\sim (\tau+\beta, \vec{x} + \beta \vec{u}_E)\,.
\eeq
As a result periodic Fourier modes now read
\beq
	\exp\left(i \omega_n \tau+i \vec{k}\cdot (\vec{x}-\vec{u}_E \tau)\right)\,,
\eeq
with $\omega_n = \frac{2\pi n}{\beta}$. In position space the action of our Models is the same as before. But the momentum-space collective field theory now reads the same as at zero velocity but with one modification. The action is
\begin{align}
\begin{split}
	S &= (N-1)\ln \text{det}(i \tilde{\omega}_n \Delta + \Sigma)  - \mu \int Dk \, G(-k,k)
	\\
	&\quad  -\int Dk_1 Dk_2 \,\Sigma(k_1,k_2)G(k_1,k_2)
	\\
	& \qquad + \int Dk_1 Dk_2 Dk_3 \,G(k_1,k_3)G(k_2,k_4) V_4(k_i)
	\\
	 & \qquad \qquad  + \int Dk (i\tilde{\omega}_n + \Sigma(-k,k)) |\tilde{\sigma}(k)|^2\,,
\end{split}
\end{align}
where
\beq
	\tilde{\omega}_n = \omega_n -  \vec{u}_E\cdot \vec{k}\,.
\eeq
Making a translation-invariant ansatz for $(G,\Sigma)$ and that the condensate is a plane wave, $\langle \sigma \rangle = \sigma e^{i \vec{\kappa}\cdot (\vec{x}-\vec{u}_E\tau)}$ with $\vec{\kappa}$ along the velocity, the Dyson equations follow and read as before, again with the shift $\omega \to \tilde{\omega}$. In particular the self-energy is frequency-independent and we end up with a single Dyson equation for $\Sigma$,
\begin{align}
\begin{split}
	\Sigma(\vec{k}) & = \int \frac{d^dk'}{(2\pi)^d} NV_4(-k,-k',k,k') 
	\\
	& \qquad \times \coth\left( \frac{\beta (\Sigma(\vec{k}')-i \vec{u}_E\cdot \vec{k}')}{2}\right)
	\\
	&\qquad \qquad  - \mu + \frac{2|\sigma|^2}{N}V_4(-k,-\kappa,k,\kappa)\,,
\end{split}
\end{align}
along with $\sigma (\Sigma(\kappa)-i \vec{u}_E \cdot \vec{\kappa}) = 0$.

Let us consider Model 1 for definiteness, where $NV_4(-k,-k',k,k') = \text{Re}(\lambda)|\vec{k}-\vec{k}'|^2 + \lambda_4$. Since the interaction is a quadratic polynomial in the external momentum $\vec{k}$, $\Sigma$ will be too. Decomposing $\vec{k}$ into a vector $\vec{k}_T$ orthogonal to $\vec{u}_E$ and a component $k_{||}$ of $\vec{k}$ parallel to $\vec{u}_E$, we may pick a rotationally-invariant ansatz for $\Sigma$ in the directions transverse to the velocity,
\beq
	\Sigma(\vec{k}) = a_1 |\vec{k}_{\perp}|^2 + b_1 k_{||}^2 + b_0 k_{||} + a_0\,.
\eeq
Assuming that $a_1$ and $b_1$ are both nonzero, regulating with heavy partner fields, and performing the momentum integrals, we find the Dyson equations
\begin{align}
\begin{split}
	a_1 = b_1 & = \frac{2}{(4\pi \beta)^{\frac{d}{2}}}\frac{\text{Re}(\lambda)}{a_1^{\frac{d-1}{2}}b_1^{\frac{1}{2}}}\text{Li}_{\frac{d}{2}}\left( \zeta\right)+\frac{2|\sigma|^2}{N}\text{Re}(\lambda)\,,
	\\
	b_0 &= \frac{2}{(4\pi \beta)^{\frac{d}{2}}}\frac{\text{Re}(\lambda)}{a_1^{\frac{d-1}{2}}b_1^{\frac{1}{2}}}\frac{b_0-iu_E}{b_1} \text{Li}_{\frac{d}{2}}\left( \zeta\right)
	\\
	& \qquad -\frac{4|\sigma|^2}{N}\text{Re}(\lambda)\kappa\,,
\end{split}
\end{align}
along with
\begin{align}
	\nonumber
	a_0 & = \frac{2}{(4\pi \beta)^{\frac{d}{2}}}\frac{1}{a_1^{\frac{d-1}{2}}b_1^{\frac{1}{2}}}\Big( \frac{\text{Re}(\lambda)}{2\beta}\Big( \frac{d-1}{a_1} + \frac{1}{b_1}\Big)\text{Li}_{\frac{d+2}{2}}(\zeta)
	\\
	\nonumber
	& \qquad \qquad \qquad  + \Big( \text{Re}(\lambda)\frac{(b_0-iu_E)^2}{4b_1}+\lambda_4\Big)\text{Li}_{\frac{d}{2}}(\zeta)\Big)
	\\
	& \qquad\qquad \qquad  -\mu +  \frac{2|\sigma|^2}{N}(\text{Re}(\lambda)\kappa^2+\lambda_4)\,,
\end{align}
where
\beq
	\zeta = \exp\left(-\beta \left( a_0 - \frac{(b_0-iu_E)^2}{4b_1}\right)\right)\,.
\eeq

Consider a $U(N)$-symmetric phase with $\sigma = 0$. Then the Dyson equation for $b_0$ can be rewritten as
\beq
	b_0 = b_0 - i u_E\,,
\eeq
which has no solution. We conclude that in an unbroken phase we must have $a_1=b_1=0$. In an unbroken phase the charge density is given by $\rho = 2\text{Re}(\lambda)a_1$, and so we see that turning on a velocity sets the charge of the equilibrium state to vanish. Unfortunately, once we set $a_1=b_1=0$, the regulated momentum integrals at hand do not obviously converge, and we have to revisit the definition of the model. One promising way forward is to turn on a small amount of dipole-breaking in the form of an ordinary spatial kinetic term $m^2 \partial_i \bar{\phi}^a\partial_i \phi^a$, solve the ensuing Dyson equations for that model, and then study the unbroken limit. Another is to simply put the model on a lattice as in the last Appendix.

Finally, we note that there is no glaring contradiction in the Dyson equations for a broken phase with $\sigma \neq 0$. A more careful analysis of the equations is required to see whether solutions of that sort exist or not.

%--------------------------------------
\section{An effective Goldstone description}
%--------------------------------------
In this Appendix we construct and study Goldstone effective descriptions for the $p$-wave and $s$-wave dipole superfluids uncovered in this work. The field content and symmetries of the description of the $p$-wave phase have already appeared in~\cite{Lake:2022ico}, although we arrived at it independently, while the description of the $s$-wave phase appeared long ago in~\cite{Pretko_2018}.

%----------------------------------------------
\subsection{$p$-wave phase}
%----------------------------------------------

In the high-temperature $p$-wave phase we have broken dipole symmetry while the global $U(N)$ symmetry remains unbroken. We then expect the low-energy description to be given in terms of a field $\vec{\pPsi}$ valued in the space of dipole transformations $\vec{d}$. This field is inert under $U(N)$ rotations and transforms as a vector under rotations, but shifts under dipole transformations,
\beq
	\vec{\pPsi} \to \vec{\pPsi} - \vec{d}\,.
\eeq
In the continuum $\pPsi_i$ is then non-compact, while in a lattice model $\vec{\pPsi}$ is valued in the dual lattice, the Brillouin zone. For a large but finite lattice the $\pPsi_i$ are discrete, and so in this Appendix we instead stick to models where $\pPsi$ is continuous, i.e. either we work on an infinite lattice or in the continuum. It is straightforward to write down effective theories of $\vec{\pPsi}$ invariant under $U(N)$ and dipole transformations, since the dipole symmetry implies that $\pPsi_i$ must be derivatively coupled. We then have
\begin{align}
\begin{split}
	S_{\rm p-wave} = \int d\tau d^dx \Big( &\frac{c_1}{2}\partial_{\tau} \pPsi_i \partial_{\tau} \pPsi_i + \frac{c_2}{2} \partial_i \pPsi_j \partial_i \pPsi_j
	\\
	& + \frac{c_3}{2} (\vec{\nabla} \cdot \vec{\pPsi})^2 + O(\partial^4)\Big)\,,
\end{split}
\end{align}
for low-energy constants $c_i$. Note that this effective description has a scaling symmetry in which time and space scale the same way. The various corrections have at least four derivatives, and so we have a Gaussian fixed point with various irrelevant corrections. 

Any plausible real-world candidate that makes contact with our models will have at best a very good approximate rather than an exact dipole symmetry. Note that it is very easy to write down effective theories of $\pPsi_i$ with small dipole breaking. In the continuum the $\pPsi_i$ then become pseudo-Goldstone bosons and the action includes mass terms $\sim m^2 \pPsi_i^2$, while on an infinite lattice the mass terms must be consistent with the periodicity of $\pPsi$, like $-\sum_i\frac{m^2}{a^2} \cos\left( a \pPsi_i\right)$ on a cubic lattice. 

At finite temperature the nonzero Matsubara modes of $\pPsi_i$ have an effective mass of order the temperature, and we expect there to be other degrees of freedom at that scale. Integrating out those modes, the finite-temperature effective description at long wavelengths is really only of the Matsubara zero mode of $\pPsi_i$, so that the time-derivative term in the action above drops out and we are left with
\beq
	S_{\rm p-wave} = \beta \int d^dx \Big(  \frac{\tilde{c}_2}{2}\partial_i \pPsi_j \partial_i \pPsi_j + \frac{\tilde{c}_3}{2} (\vec{\nabla}\cdot \vec{\pPsi})^2+ O(\partial^4)\Big)\,.
\eeq

Lattice regularized theories have compact $\pPsi_i$ and therefore have operators charged under dipole number in the deep infrared. On a cubic lattice so that $\pPsi_i \sim \pPsi_i + \frac{2\pi}{a}$, the basic charged operators are $e^{i a \pPsi_i}$. We may understand these operators more microscopically in a lattice model with a charged scalar field $\phi$. In that setting, consider the scalar bilocal $G_{x,y} = \bar{\phi}(x)\phi(y)$. Now let $x$ and $y$ be at the same time, but with $x$ one lattice site away from $y$ in the $i^{\rm th}$ direction. Then employing a polar decomposition $\phi = v e^{i\varphi}$, we have
\beq
\label{E:microscopicPsi}
	\frac{G_{x,y}}{\sqrt{G_{x,x}G_{y,y}}} = \exp\left( -i(\varphi(x)-\varphi(y)\right)\equiv  \exp\left(-i a \pPsi_i(y)\right)\,.
\eeq
In a $U(N)$-symmetric but dipole-breaking phase, the order parameter $\phi$ has no expectation value and so $\varphi$ is not a sensible low-energy degree of freedom, but $G_{x,x}$ and $G_{x,y}$ certainly do have expectation values, making the gradient in $\varphi$ well-defined in the low-energy description. This leads to a collective field $\pPsi_i$, which by this identification has periodicity $\frac{2\pi}{a}$.

Incidentally, this identification of $\pPsi$ informs us that $\pPsi$ is a periodic field even on a finite lattice where the dipole group is finite.

With this low energy description in hand we may assess whether or not the dipole symmetry is restored by infrared fluctuations. We refer the reader to the discussion in \cite{Distler:2021qzc} for more detail. By the same methods used there, we find that long-wavelength fluctuations of the $\pPsi_i$ restore dipole symmetry at finite temperature in $d\leq 2$, but not in $d>2$. In our large $N$ models these infrared fluctuations are a $1/N$ effect and therefore invisible in our leading large $N$ approximations. We then expect that at finite but large $N$ our $d=2$ models have dipole quasiorder, analogous to the quasiorder of two-dimensional vector models~\cite{Witten:1978qu}.

%----------------------------------------------
\subsection{$s$-wave phase}
%----------------------------------------------

Let us ignore the complications that come with the fact that we really have a $U(N)$ global symmetry that is broken to $U(N-1)$, and instead consider a scenario where the global symmetry $U(1)$ is completely broken. Now we need to include an additional compact field $\varphi$ for the broken $U(1)$, where under $U(1)$ and dipole transformations $\varphi$ shifts as
\beq
	\varphi \rightarrow \varphi - \alpha - d_i x^i\,.
\eeq
The lowest order effective action for $\varphi$ is
\begin{equation}
    S_{\rm s-wave}  =\int d\tau d^dx \left(\frac{b_1}{2}(\partial_\tau \varphi)^2 + \frac{b_2}{2}\left(\nabla^2 \varphi \right)^2+O(\partial^4)\right)\,.
\end{equation}

In fact, we could write down effective theories of $\varphi$ and $\pPsi_i$ as well, in which case the symmetries allow for the quadratic term $\left(\pPsi_i - \partial_i\varphi \right)^2$. However, at low momentum this term effectively gaps $\pPsi_i$, soldering it to $\partial_i \varphi$, and so we are only left with an effective description of $\varphi$. Further, at finite temperature, the Matsubara modes of $\varphi$ are also gapped and at low energy we only have the Matsubara zero mode of $\varphi$.

It is also straightforward to incorporate the effect of dipole breaking in the effective description of $\varphi$. In the continuum the lowest-order breaking term is $|\vec{\nabla}\varphi|^2$, while global symmetry forbids mass terms like $-m^2 \cos(\varphi)$. 

The basic charged operator built from $\varphi$ is the phase $e^{i\varphi}$, which carries charge both under $U(1)$ and dipole. Products of charges and anticharges can be arranged to carry vanishing net charge but nonzero dipole number. To assess whether long-wavelength fluctuations restore the $U(1)$ and dipole symmetries, we can use the methods of~\cite{Distler:2021qzc} to find that the symmetry is restored at finite temperature in $d\leq 4$. These fluctuations are a $1/N$ effect, and so are invisible in our large $N$ analysis. Our large $N$ models in $d=3,4$ then have a high-temperature phase with unbroken $U(1)$ but broken dipole number, and a low-temperature phase in which the $U(1)$ is quasiordered but the $SU(N)$ is broken to $SU(N-1)$.

%----------------------------------------------
\subsection{Lattice non-decoupling}
%----------------------------------------------
We turn our attention to the non-decoupling of an underlying lattice in the low-energy Goldstone effective action. See~\cite{Seiberg:2020bhn,Seiberg:2020wsg} for analyses in the context of Goldstone models of condensed phases in models with subsystem symmetry, and~\cite{Gorantla:2022eem} for a discussion of non-decoupling in putative condensed phases of 1+1-dimensional models with dipole symmetry.

Here we initiate a discussion of the subtleties that can arise, focusing largely on the $p$-wave phase and deferring a more thorough investigation to the future.

As we have seen earlier in the Appendix, dipole transformations are valued in the space of momenta, the dual lattice $\widetilde{\Gamma}$. So the dipole group is non-compact in the continuum, compact but continuous on an infinite lattice, and compact but discrete on a finite lattice. This leads to a number of differences between the continuum Goldstone models, and their lattice formulations, which in principle can be obtained in our large $N$ models by a careful analysis of the four-point function.

There are three differences we point out here between the models in the continuum, on an infinite lattice, and on a finite lattice:
\begin{enumerate}
	\item The dipole Goldstones $\pPsi_i$ are non-compact in the continuum and compact with a periodicity $\sim \frac{2\pi}{a}$ on the lattice. (On a cubic lattice this is the periodicity, but in general the $\pPsi_i$ are valued in the Brillouin zone.) Similarly, for the low-temperature theory of a scalar $\varphi$, the zero modes corresponding to dipole breaking are non-compact in the continuum, but compact on a lattice. The finite volume path integral is sensitive to the compactness or non-compactness of the Goldstone fields, as the integral over the zero modes produces a factor of the (normalized) dipole symmetry group. This is infinite in the continuum, while on a lattice it goes as $VV_{\rm BZ} = N_{\rm sites}$. 
	\item On a lattice the compactness of the dipole transformations implies the existence of winding modes of the Goldstones around the thermal or spatial circles. For example, on a lattice, we must sum over configurations where the $\pPsi_i$ wind around the thermal circle, which for a cubic lattice read
	\beq
		\pPsi_i = \frac{n_i \tau}{\beta} \frac{2\pi}{a} + \delta \Psi_i\,,
	\eeq
	where the $\delta \Psi_i$ are periodic around the circle. On an infinite lattice these configurations have infinite action, but on a finite lattice they have an action $\sim \frac{c_1|\vec{n}|^2 V}{a^2\beta}$, which implies the existence of low-energy states of energy $\sim \frac{a^2}{c_1V}|\vec{m}|^2$ for some integer-valued vector $\vec{m}$. These states are present and exchanged in physical processes for the theory on a finite lattice, but not on an infinite lattice (where these winding modes are absent) or in the continuum (where there are no winding modes at all).
	\item Finally the dipole symmetry is finite on a finite lattice. In~\eqref{E:microscopicPsi} we have given a lattice definition of the dipole Goldstone $\pPsi_i$ which is valued in the Brillouin zone even on a finite lattice, but only a discrete set of values for the $\pPsi_i$ would then be saddle points of the model. That is, we expect the Goldstone model on a finite lattice to have potential terms $\sim- \sum_i \cos\left( L_i a \pPsi_i\right)$ where $L_i$ is the number of sites in the $i^{\rm th}$ direction, so that only $\pPsi_i = \frac{2\pi}{a} \frac{n}{L_i}$ are genuine saddles of the Goldstone model. That being said, continuity of the large $N$ effective action in the limit of small lattice spacing and large number of sites suggests that these potential terms are suppressed by powers of $L_i$ in the large volume limit.
\end{enumerate}

To match any candidate real-world system there will inevitably be some small amount of dipole breaking. As long as that breaking is macroscopic, i.e. not suppressed by factors of the lattice spacing, we expect these subtleties concerning the Goldstone modes to be irrelevant, i.e. these Goldstone models ought to be adequate for the practical purpose of predicting macroscopic properties of systems with approximate dipole symmetry.

%----------------------------------------------
\subsection{Coupling to background}
%----------------------------------------------

It was shown in~\cite{jensen2021} (see also~\cite{Bidussi:2021nmp}) how to couple models with a conserved dipole moment to a spacetime background. Provided that the sources we turn on are time-independent and slowly varying in space, we can understand how these sources to appear in the long-wavelength and zero-frequency Goldstone models discussed here. The effective action so constructed describes equilibrium configurations of the ``fracton hydrodynamics'' to appear in~\cite{Hydro}, much like how finite-temperature effective actions for an ordinary Goldstone mode describe hydrostatic equilibria of superfluid hydrodynamics (see \cite{Banerjee:2012iz,Jensen:2012jh,Bhattacharyya:2012xi}). 

A full discussion of the $p$-wave and $s$-wave effective actions in background and the matching to hydrodynamics will appear in~\cite{Hydro}. Here though we study a simple feature, namely the coupling of the flat space Goldstone models to a source $A_{ij}$ for dipole current and a source $A_{\mu}$ for the ordinary current, with
\beq
	\delta S = \int d\tau d^dx \left( J^{\mu} \delta A_{\mu}+ \frac{1}{2} J^{ij}\delta A_{ij}\right)\,.
\eeq
In coupling to background we mandate invariance under $U(1)$ gauge transformations,
\beq
	A_{\mu} \to A_{\mu} + \partial_{\mu} \Lambda\,,
\eeq
along with invariance under local dipole transformations,
\beq
	A_i \to A_i + \psi_i\,, \qquad A_{ij} \to A_{ij} + \partial_i \psi_j + \partial_j \psi_i\,.
\eeq
$U(1)$ invariance implies that the charge current is conserved, $\partial_{\mu} J^{\mu}=0$, while the dipole symmetry implies 
\beq
	J^i = \partial_j J^{ij}\,,
\eeq
so that putting the two together implies
\beq
	\dot{\rho} + \partial_i \partial_j J^{ij} = 0\,,
\eeq
with $J^0=\rho$.

We endow the Goldstone fields with transformation laws 
\begin{align}
\begin{split}
	\pPsi_i& \to \pPsi_i + \psi_i\,,
	\\
	\varphi & \to \varphi - \Lambda \,.
\end{split}
\end{align}
At low temperature, if we include both $\varphi$ and $\pPsi$ in the effective description, we can define twisted versions of $A_{\mu}$ and $A_{ij}$ invariant under $U(1)$ and dipole transformations,
\begin{align}
\begin{split}
	\widetilde{A}_{\tau} & = A_{\tau} + \partial_{\tau} \varphi\,,
	\\
	\widetilde{A}_i & = A_i + \partial_i \varphi - \pPsi_i\,,
	\\
	\widetilde{A}_{ij} & = A_{ij} - \partial_i \pPsi_j - \partial_j \pPsi_i\,.
\end{split}
\end{align}
At high temperature where we only have $\pPsi$, $\widetilde{A}_{ij}$ is an invariant object while $\widetilde{A}_{\mu}$ is dipole invariant but transforms as a connection under $U(1)$ transformations. 

With these building blocks in hand we can write long-wavelength actions Goldstones in the presence of time-independent background fields $A_{\mu}$ and $A_{ij}$. At high temperature and only including terms that remain when setting background fields to vanish we have
\beq
	S_{\rm p-wave} = \beta \int d^dx \left(  \frac{a_1}{4}\widetilde{F}_{ij}^2 + \frac{a_2}{4} \widetilde{A}_{\{ij\}}^2 + \frac{a_3}{4} (\widetilde{A}_{ii})^2+O(\partial^4)\right)\,,
\eeq
where the $a_i$ are low-energy constants that depend on control parameters. At low temperature in the $s$-wave phase we have
\begin{align}
\begin{split}
	S_{\rm s-wave} & = \beta \int d^dx \Big( \frac{b_0}{2} \widetilde{A}_i^2 + \frac{b_1}{2} (\partial_i \widetilde{A}_i)^2 + \frac{b_2}{4} \widetilde{F}_{ij}^2 
	\\
	& \qquad + \frac{b_3}{4} \widetilde{A}_{\{ij\}}^2 + \frac{b_4}{2} \widetilde{A}_{ii}^2 + O(\partial^4)\Big)\,,
\end{split}
\end{align}
where the $b_i$ are low-energy constants.

The dipole and vector currents follow by variation. At high temperature we have
\begin{align}
\begin{split}
	J^{ij} &= a_2 \widetilde{A}^{\{ij\}} + a_3 \delta^{ij} \widetilde{A}_{kk} + O(\partial^3)\,,
	\\
	J^i &= a_1 \partial_j \widetilde{F}^{ij}+O(\partial^4)\,,
\end{split}
\end{align}
and the equation of motion for the dipole Goldstones $\pPsi_i$ is simply the dipole Ward identity $J^i = \partial_j J^{ij}$. In the low-temperature $s$-wave phase the term $b_0$ acts as a mass term for the $\pPsi_i$, and upon integrating them out we have a theory of the scalar Goldstone $\varphi$. After integrating out the $\pPsi_i$, the ensuing currents $J^i$ and $J^{ij}$ automatically satisfy the dipole Ward identity, and the equation of motion for $\varphi$ is current conservation $\partial_{\mu} J^{\mu} = 0$.

These effective actions compute low-momentum, zero frequency response functions. At tree level one obtains the propagators for the Goldstones from the effective action and then computes e.g. $\langle J^{ij} J^{kl}\rangle$ via Wick contractions. By coupling to a spacetime background one can also compute low-momentum, zero frequency response functions of the energy current, momentum current, and stress tensor. We defer a more complete analysis of these response functions to~\cite{Hydro}.

%--------------------------------------
\section{High and low temperature limits} \label{app:limits}
%--------------------------------------
In this Appendix we will derive the high and low temperature limits of the solutions to the Dyson equations for both Model 1 and Model 2. 

%------------------------------------------
\subsection{Model 1}
%------------------------------------------

The Dyson equations~\eqref{E:dyson} of Model 1 are relatively simple, becoming~\eqref{eq:dyson_mod1_N} in the normal phase and~\eqref{eq:dyson_mod1_C} in the condensed phase. Let us recapitulate those equations. For the normal phase we have
\begin{align*}
  a_0 & =  -\mu +  \frac{d}{2\beta} \frac{\text{Li}_{\frac{d+2}{2}}\left( e^{-\beta a_0}\right)}{\text{Li}_{\frac{d}{2}}\left( e^{-\beta a_0}\right)}  + \frac{\lambda_4}{\text{Re}(\lambda)} a_1 \,,
  	 \\
	a_1 & = \frac{2 \text{Re}(\lambda)}{(4\pi \beta a_1)^{\frac{d}{2}}}\text{Li}_{\frac{d}{2}}\left(e^{-\beta a_0}\right) \,,
\end{align*}
and in the condensed phase we have
\begin{align*}
	 \mu & = \frac{\lambda_4}{\text{Re}(\lambda)} a_1  +  \frac{d~ \text{Re}(\lambda)}{(4\pi \beta a_1)^{\frac{d}{2}}} \frac{1}{\beta a_1} \zeta\left(\frac{d+2}{2}\right)\,,
	 \\
    \frac{|\sigma|^2}{N} &= \frac{a_1}{2 \text{Re}(\lambda)}  -  \frac{1}{(4\pi \beta a_1)^{\frac{d}{2}}}\zeta\left(\frac{d}{2}\right) \,.
\end{align*}

Let us first consider $d=2$, for which the Dyson equations of the condensed phase diverge and so in fact a condensed phase does not exist. At low temperature and fixed positive chemical potential we find a solution where $\beta a_0\to 0$ even as $\beta\to\infty$. We then have
\beq
    a_1 = \frac{\text{Re}(\lambda)\mu}{\lambda_4} + O(\beta^{-1})\,,
\eeq
 and $\beta a_0$ satisfies
\begin{equation}
    -\ln(\beta a_0) = \frac{2\pi \beta \text{Re}(\lambda) \mu^2}{ \lambda_4^2}  + O(\beta^0)\,,
\end{equation}
so
\begin{equation}
    a_0 \sim \frac{e^{- \beta \frac{2\pi \text{Re}(\lambda) \mu^2}{ \lambda_4^2}}}{\beta} \,.
\end{equation}
As the charge density of Model 1 is $\rho = 2 \text{Re}(\lambda)a_1$, we see that the low-temperature limit of Model 1 at $\mu>0$ is a state of compressible matter with $\frac{\partial \rho}{\partial \mu} = \frac{2\text{Re}(\lambda)^2}{\lambda_4}$ and an effective mass $a_0$ that tends to zero exponentially fast.

The low temperature, $\mu<0$ limit of the normal phase is basically the same in $d=2$ as in higher dimension. Using the asymptotic form of the Polylogarithm
\begin{equation}
    \text{Li}_a(x) = x + 2^{-a} x^2 + O(x^3)\,,
\end{equation}
we see that when $\mu < 0 $ the solutions to the Dyson equations in this limit are
\begin{equation}
\begin{split}
    a_0 &= -\mu + \frac{d}{2\beta} + O(\beta^{-2}),\\
    a_1 &= \left(\frac{2 \text{Re}(\lambda)}{(4\pi \beta)^{d/2}}\right)^{2/(d+2)} e^{(2\beta \mu- d)/(d+2)} + \ldots
\end{split}
\end{equation}
and $\sigma = 0$. In this limit the Green's function asymptotes to a zero temperature form
\beq
	\lim_{T\to 0}	G= \frac{1}{i\omega - \mu}\,,
\eeq
which notably displays no dipole symmetry breaking. Since the charge density is $\rho = 2 \text{Re}(\lambda)a_1$ and $a_1\to 0$, we see that there is also no charge density in that limit. So in the limit of zero temperature and $\mu < 0$ we find a theory of nothing. At any nonzero temperature though we have a small but nonzero charge density with an exponentially suppressed compressibility $\frac{\partial \rho}{\partial \mu}\sim e^{\frac{2\beta\mu}{d+2}}$.

At low temperature, $\mu>0$, and $d>2$ there are no solutions to the Dyson equations of the normal phase. The Model condenses to a $U(N)$-breaking phase at low temperature and positive $\mu$ which we presently discuss. Expanding the Dyson equations of the condensed phase at low temperature and positive fixed $\mu$ we find a solution
\begin{equation}
    \begin{split}
        a_1 &= \frac{\text{Re}(\lambda) \mu}{\lambda_4} - \frac{d  (\text{Re}(\lambda))^{(2 - d)/2}\zeta\left(\frac{d+2}{2}\right)}{(4\pi)^{d/2} (\beta \mu/\lambda_4)^{(d+2)/2}}  + O(\beta^{-d-2}),\\
        \frac{\sigma^2}{N} &= \frac{\mu}{2\lambda_4} - \frac{\zeta\left(\frac{d}{2} \right)}{(4\pi\text{Re}(\lambda) \mu \beta/\lambda_4 )^{d/2} } + O(\beta^{-(d+2)/2}),
    \end{split}
\end{equation}
and $a_0 = 0$. At zero temperature the condensate is precisely what we get for the minimum of the classical potential $-\mu |\sigma|^2 +\frac{ \lambda_4}{N} |\sigma|^4$ as expected from a large $N$ vector model. We also note that the parameter $a_1$ assumes the simple form $\frac{\text{Re}(\lambda)\mu}{\lambda_4}$ at zero temperature. This is exactly what we would find at one-loop level were we to treat the quartic interaction $\text{Re}(\lambda)$ as a perturbative interaction when expanding around the minimum of the classical potential. In this sense, the condensed zero temperature phase is one-loop exact in $\lambda$ to leading order in large $N$.

There are no solutions to the Dyson equations of the condensed phase when $\mu<0$. 

Now let us study the high temperature limit at fixed chemical potential. The Dyson equations of the condensed phase have no solutions in this regime, so we consider the normal phase. This task is a bit more involved as it is not clear how $a_0$ and $a_1$ should scale with $\beta$ from the Dyson equations. Assuming $\beta a_0$ approaches some finite value (which might be zero) then $a_1$ scales like $\beta^{-d/(d+2)}$. We then have $\beta a_1 \rightarrow 0$ in this limit, leading to the self consistent equation for $\beta a_0$:
\begin{equation}
    \beta a_0 \approx \frac{d}{2}  \frac{\text{Li}_{\frac{d+2}{2}}\left( e^{-\beta a_0}\right)}{\text{Li}_{\frac{d}{2}}\left( e^{-\beta a_0}\right)}.
\end{equation}
Thus in this limit we have that
\begin{equation}
    a_0 \approx \frac{a_0^*}{\beta}, \quad 
    a_1 \approx \left[\frac{2 \text{Re}(\lambda) }{(4\pi)^{d/2}}\text{Li}_{\frac{d}{2}}\left(e^{- a_0^*}\right) \right]^{2/(d+2)} \beta^{-d/(d+2)},
\end{equation}
where $a_0^*$ is the solution to the equation
\begin{equation}
    a_0^* = \frac{d}{2}  \frac{\text{Li}_{\frac{d+2}{2}}\left( e^{- a_0^*}\right)}{\text{Li}_{\frac{d}{2}}\left( e^{- a_0^*}\right)} .
\end{equation}
Note that the self-energy depends non-analytically on $\text{Re}(\lambda)$ in this limit. 

We can numerically solve the Dyson equations for any temperature and chemical potential. When we do so we find that our numerical solutions nicely match the low- and high-temperature limits obtained here.

%------------------------------------------
\subsection{Model 2}
%------------------------------------------

The Dyson equations for Model 2 are quite lengthy and appear in~\eqref{eq:SD_mod2}. We endeavor to solve them in the low and high-temperature limits.

We begin at low temperature in $d>2$ with $\text{Re}(\gamma)>0$. At negative chemical potential, the only solutions are those describing a normal phase, and in that regime the integral over $k'$ in those equations vanishes in the low-temperature limit. This leads to a zero temperature solution
\beq
	a_0 =-\mu\,, \qquad a_1=a_2=0\,,
\eeq
i.e. a zero-temperature Green's function
\beq
	\lim_{T\to 0}G = \frac{1}{i\omega-\mu}\,,
\eeq
describing a theory of nothing. The low-temperature corrections to this solution are exponentially small, going as powers of $e^{\beta \mu}$. On the other hand, for $\mu>0$, the only solutions are condensed. They behave as
\begin{equation}
    \begin{split}
        a_1 &= \frac{2 \text{Re}(\gamma) \lambda_S \mu}{\lambda_S|\gamma|^2 + \lambda_4} + O(\beta^{-\frac{d+2}{2}}) \,,
         \\
        a_2 &= \frac{ \lambda_S \mu}{\lambda_S|\gamma|^2 + \lambda_4} + O(\beta^{-\frac{d+2}{2}}) \,,
        \\
        \frac{|\sigma|^2}{N} &= \frac{ \mu}{2\left(\lambda_S|\gamma|^2 + \lambda_4 \right)} + O(\beta^{-\frac{d}{2}})\,.
    \end{split}
\end{equation}
The denominator in these expressions is the full quartic coupling $\lambda_S |\gamma|^2+\lambda_4$ appearing in the scalar potential. Indeed, as in Model 1, there is a sense in which this solution is one-loop exact in the quartic couplings with spatial derivatives, to leading order in large $N$. The effective classical potential is $V = -\mu |\sigma|^2 + (\lambda_S |\gamma|^2+\lambda_4)|\sigma|^4/N$, which is minimized precisely at the condensate above. Expanding in fluctuations around that minimum, i.e. treating $\lambda_S$ and $\gamma$ perturbatively, the results for $a_1$ and $a_2$ come from one-loop diagrams. 

The low-temperature behavior for $\text{Re}(\gamma)<0$ is quite different than for $\text{Re}(\gamma)>0$. For $\text{Re}(\gamma)<0$ and $\mu>0$, $a_1$ becomes negative at low temperature, implying that $a_0 > 0$ in order for the integrals in the Dyson equations~\eqref{eq:SD_mod2} to converge, and so $\sigma=0$. If this is the case then we can write the self energy $\Sigma(k)$ as $\Sigma(k) =  A_0 + a_2 \left(k^2-k_0^2 \right)^2,$ where $k_0^2$ and $A_0$ are related to $a_0$ and $a_1$ by
\begin{equation}
    a_0 = A_0 + a_2 k_0^4, \quad a_1 = - 2 a_2 k_0^2.
\end{equation}
Then in the large $\beta$ limit we can evaluate the integrals in \eqref{eq:SD_mod2} using a saddle point approximation around the saddle $k = k_0$. As $a_2$ is finite in this limit, the last equation in \eqref{eq:SD_mod2} implies that $A_0 \rightarrow 0$. The remaining two equations reduce to the linear set
\begin{align}
\nonumber
    \mu  &= a_2 \left(|\gamma|^2 + \frac{\lambda_4}{\lambda_S} - \frac{2d~\text{Re}(\gamma) (k_0^2 + \text{Re}(\gamma))}{d+2} \right)\,,
      \\
     - 2 a_2k_0^2 &=  2a_2\left( \text{Re}(\gamma) + \frac{(d+2)}{d} k_0^2\right)\,,
\end{align}
up to corrections of the order $\beta^{-1}$. This gives us the low temperature solution
\begin{equation}
\begin{split}
    k_0^2 &= - \frac{d\text{Re}(\gamma)}{2(d+1)} ,\\
    a_2   &= \mu \left(|\gamma|^2 + \frac{\lambda_4}{\lambda_S}  - \frac{d}{d+1} \text{Re}(\gamma)^2\right)^{-1} .
\end{split}
\end{equation}
Notice that $a_2$ diverges for sufficiently large and negative $\text{Re}(\gamma)$, resulting in a condensate at some finite nonzero value of momentum $|k| = k_0$, when $\text{Re}(\gamma) = - \sqrt{\frac{(d+1) (|\gamma|^2 + \lambda_4/\lambda_S)}{d}}$. However this occurs in the regime where the forward scattering is no longer strictly positive, and so is not physically relevant. 

Now consider the high temperature limit at fixed chemical potential. Our numerical solutions are all in the normal phase in this regime, and they indicate that $\beta a_0, \beta a_1\gg \beta a_2 \rightarrow 0$. Numerically we also see that $\beta a_0$ tends to a finite value as $\beta\to 0$. Thus in order to find an approximate solution we drop the coefficient $a_2$ in the Dyson equations~\eqref{eq:SD_mod2}, and show that this approximation is self-consistent. Under it we can evaluate the integrals in \eqref{eq:SD_mod2}, resulting in 
\begin{widetext} 
\begin{equation} \label{eq:sd_mod2_T_inf}
    \begin{aligned}
     a_0 &=  \left(|\gamma|^2+\frac{\lambda_4}{\lambda_S}\right)a_2 
     + \frac{d}{d+2}\text{Re}(\gamma)\left(a_1 - \text{Re}(\gamma) a_2\right) +\lambda_S (\beta a_1)^{-\frac{d+4}{2}}  \Gamma\left(\frac{d+4}{2}\right)\text{Li}_{\frac{d+4}{2}}\left(e^{-\beta a_0}\right)\, ,
      \\
     a_1 &=  2\text{Re}(\gamma) a_2 + \lambda_S  \frac{2(d+2)}{d} (\beta a_1)^{-\frac{d+2}{2}}\Gamma\left(\frac{d+2}{2}\right)\text{Li}_{\frac{d+2}{2}}\left(e^{-\beta a_0}\right) \,,
     \\
     a_2 &=  \lambda_S  (\beta a_1)^{-\frac{d}{2}}\Gamma\left(\frac{d}{2}\right)\text{Li}_{\frac{d}{2}}\left(e^{-\beta a_0}\right)\,.
    \end{aligned}
\end{equation}
\end{widetext} 
These polylogarithms are bounded, and so we can combine the last two equations of \eqref{eq:sd_mod2_T_inf} to form a single equation for $a_1$,
\begin{align}
 \label{eq:sd_mod2_T_inf_2}
    a_1 &=  2\text{Re}(\gamma) \lambda_S(\beta a_1)^{-\frac{d}{2}}  \Gamma\left(\frac{d}{2}\right)\text{Li}_{\frac{d}{2}}\left(e^{-\beta a_0}\right) 
    \\
    \nonumber
    &  + \lambda_S  \frac{2(d+2)}{d} (\beta a_1)^{-\frac{d+2}{2}}\Gamma\left( \frac{d+2}{2}\right)\text{Li}_{\frac{d+2}{2}}\left(e^{-\beta a_0}\right) \,.
\end{align}
At high temperature we have $\beta a_1  \rightarrow 0$ (this is a consequence of $a_2 \rightarrow \infty$ in this limit) so we can neglect the first term in \eqref{eq:sd_mod2_T_inf_2}, resulting in the asymptotic behavior
\begin{equation} \label{eq:high_T_mod2}
    a_0 = C_0 \beta^{-1}\,, \quad 
    a_1 = C_1 \beta^{-\frac{d+2}{d+4}}\,, \quad 
    a_2 = C_2 \beta^{-\frac{d}{d+4}}\,,  
\end{equation}
where $C_0$ is the solution to the equation
\begin{equation}
        C_0 = \frac{d}{4} \frac{\text{Li}_{d/2+2}\left(e^{- C_0}\right)}{\text{Li}_{d/2+1}\left(e^{- C_0}\right)},
\end{equation}
and 
\begin{align}
        C_1 &= \left[\lambda_S (d+2) \Gamma\left(\frac{d}{2}\right)\text{Li}_{\frac{d+2}{2}}(e^{- C_0}) \right]^{\frac{2}{d+4}}\,,
        \\
        \nonumber
        C_2 &= \left[\lambda_S \Gamma\left(\frac{d}{2}\right)\right]^{\frac{4}{d+4}}\text{Li}_{\frac{d}{2}}(e^{- C_0}) \left[ (d+2)\text{Li}_{\frac{d+2}{2}}(e^{- C_0}) \right]^{-\frac{d}{d+4}}.
\end{align}

It is interesting to note that if we try to solve the high temperature limit by first dimensionally reducing the collective theory on the thermal circle, and solving the ensuing Dyson equations, we find an incorrect result, different from the one above. Evidently when it comes to the large $N$ solution, the thermal circle does not decouple at long wavelength. This is the case for both Model 1 and Model 2. However we expect fluctuations around the large $N$ solution to be well-described by an effective field theory obtained by reduction on the thermal circle.

\end{document}